\documentclass[preprint]{aastex}
\usepackage{natbib,lscape}
\usepackage{graphicx,color}
\def\red#1 {\textcolor{red}{#1}~}
\def\grn#1 {\textcolor{green}{#1}~}
\def\blue#1 {\textcolor{blue}{#1}~}

\newcommand{\etal}   {{\rm \ et~al.}}
\newcommand{\kms}  {\ifmmode{{\rm ~km\,s}^{-1}}\else{~km~s$^{-1}$}\fi}

\shortauthors{Greenhill \etal} 
\shorttitle{H$_2$O Disk Masers in AGN}

\begin{document}

\title{Discovery of Candidate H$_2$O Disk Masers in AGN and Estimations of Centripetal Accelerations}

\author{Lincoln J. Greenhill, Paul T. Kondratko,\altaffilmark{1} James M. Moran, \and Avanti Tilak}
\affil{Harvard-Smithsonian Center for Astrophysics, 60 Garden St.,
Cambridge, MA 02138 USA}

\altaffiltext{1}{Current address: Goldman Sachs, London, UK.}

\email{greenhill@cfa.harvard.edu}

\begin{abstract}
Based on spectroscopic signatures, about one-third of known H$_2$O maser sources in active galactic nuclei (AGN) are believed to arise in highly inclined accretion disks around central engines.  These ``disk maser candidates'' are of interest primarily because angular structure and rotation curves can be resolved with interferometers, enabling dynamical study. We identify five new disk maser candidates in studies with the Green Bank Telescope, bringing the total number published to 30.  We discovered two (NGC\,1320, NGC\,17) in a survey of 40 inclined active galaxies ($v_{sys}< 20000$\,km\,s$^{-1}$).  The remaining three disk maser candidates were identified in monitoring of known sources: NGC\,449, NGC\,2979, NGC\,3735.  We also confirm a previously marginal case in UGC\,4203.  For the disk maser candidates reported here, inferred rotation speeds are 130--500\,\kms. Monitoring of  three more rapidly rotating candidate disks (CG\,211, NGC\,6264, VV\,340A) has enabled measurement of likely orbital centripetal acceleration, and estimation of central masses (2--7$\times10^7$\,M$_\odot$) and mean disk radii (0.2--0.4\,pc).  Accelerations may ultimately permit estimation of distances when combined with interferometer data.  This is notable because the three  AGN are relatively distant ($10000<v_{sys}<15000$\,\kms), and fractional error in a derived Hubble constant, due to peculiar motion of the galaxies, would be small.  As signposts of highly inclined geometries at galactocentric radii of $\sim 0.1$--1\,pc, disk masers also provide robust orientation references that allow analysis of (mis)alignment between AGN and surrounding galactic stellar disks, even without extensive interferometric mapping.  We find no preference among published disk maser candidates to lie in high-inclination galaxies.  This provides independent support for conclusions that in late-type galaxies, central engine accretion disks and galactic plane orientations are not correlated.


\end{abstract}


\section{INTRODUCTION}

Extragalactic H$_2$O maser emission from active galactic nuclei (AGN) in spiral galaxies in some cases traces highly inclined disk structures at radii $\sim 0.1$--1\,pc from the central engines.  Cases established by direct interferometric imaging that resolves position-velocity structure include NGC\,4258 \citep[e.g.,][]{Miyoshi1995,Argon2007}, NGC\,1068 \citep{Greenhill1997}, NGC\,3079 (Yamauchi et al. 2004; Kondratko, Greenhill, \& Moran 2005)\nocite{Yamauchi2004, Kondratko2005},  Circinus \citep{Greenhill2003}, UGC\,3789 \citep{reid09}, and NGC\,6323 \citep{Braatz2007}. These ``disk masers''  have common spectroscopic characteristics, highly red- and blueshifted emission lines, called ``high-velocity" emission lines, that more or less evenly bracket the systemic velocity, and emission close to the systemic velocity, called ``low-velocity" emission lines. Both are defined in the frame of the rotating disk, projected along the line of sight. We use this terminology of ``high-" and ``low-" velocity lines throughout this paper. This template is also observed in other H$_2$O masers  for which imaging data are not available, and it can be a powerful diagnostic.   In total, at least 30 published masers (out of 107  known) share these  characteristics \citep[e.g.,][and references therein]{Greenhill2008}.  The high-velocity red- and blueshifted emission has been observed to be offset on the order of $10^2$--$10^3$\,\kms, among sources. We note that very long baseline interferometric (VLBI) imaging is essential to confirm the identity of disk masers. Hence, we have used the terminology ``disk maser candidates" for sources with spectroscopic identification only. We also note that VLBI imaging of some disk masers shows signs of additional maser emission arising in outflows (e.g., Circinus) and jets (e.g., NGC\,1068).

H$_2$O disk masers highlight structures oriented edge-on, or nearly so, and in close proximity to central engines.  This is plausible because large reservoirs of molecular gas along lines of sight are required to generate detectable emission \citep{eli82}. Among all H$_2$O masers in AGN, $95\%$ exhibit  X-ray absorption column densities $>10^{23}$\,cm$^{-2}$ (Greenhill et al.\ 2008)\nocite{Greenhill2008}.   Among disk masers, the fraction is nearly as high, and the distribution of obscuring column densities peaks above $10^{24}$\,cm$^{-2}$.  For thin disks, maser emission is beamed narrowly about the  midplane, into a solid angle $\ll4\pi$, as dictated by requisite velocity coherence along lines of sight \citep[e.g.,][]{Miyoshi1995}.    Geometry limits observable emission to disks that are highly inclined.  Although common association has been well recognized with optical type-2 and partly obscured type-1 objects \citep[e.g.,][]{Braatz1997b}, disk maser emission is a sharper indicator of (inclination) orientation for nuclear structure than optical classifications are.  In addition, it is a more direct indicator than, e.g., models of Fe K$\alpha$ emission from hot accretion disks \citep{nandra97}, the differential brightness of radio jets, and models of kiloparsec-scale ionization structures.  Comparisons of inclination complement analyses of position angles and naturally enable analyses of relative 3-D  orientation \citep{Clarke1998, Pringle1999, Nagar1999}.

We conducted a survey of AGN to detect new disk masers, selecting galaxies with inclined stellar disks (inclination angle $i>70^\circ$).  We detected two disk maser candidates and one unclassified maser (Section\,3.1).  We also conducted deep integrations to identify indicators of disk maser emission among previously  known sources, detecting another three  (Section\,3.2), and  monitored another three known disk maser candidates to measure centripetal acceleration due to disk rotation (Section\,3.3).  We discuss estimation of black hole masses and disk radii in Section\,4 as well as the distribution of inclinations among disk maser candidate hosts and  misalignment between central engines and galactic stellar disks.  

\section{OBSERVATIONS}

The survey was conducted during 2004--2005 with the Green Bank Telescope (GBT) of the NRAO,\footnote{The
National Radio Astronomy Observatory is operated by Associated Universities
Inc., under cooperative agreement with the National Science Foundation.} using two 200-MHz IFs configured to overlap by 50~MHz, in each polarization. We tuned the center of the resulting 350-MHz instantaneous bandwidth to the systemic velocity of each target
source (350~MHz $\sim 4880$\kms\ for a representative
recessional velocity of 5000\kms, assuming the optical definition of
Doppler shift). The spectrometer delivered a channel spacing of 24.4\,kHz.  
The dedicated observation of known H$_2$O maser sources was conducted
between 2003 and 2006 with the same setup, except for galaxy CG\,211 (2003 December 08), where the two IFs were configured to overlap by 20\,MHz and the IF at the lower velocity was centered on the systemic velocity of the galaxy. 

To obtain total-power spectra, we nodded the telescope by $3'$
every 2\,minutes between two positions on the sky so that each target was always
present in the on-axis or off-axis beam. System temperatures ($T_{\rm sys}$)
were measured against a calibrated noise source injected at the receiver and
ranged from $23$ to $75$\,K depending on elevation and weather conditions. To
calibrate the spectra to flux density units, we used the gain curve obtained by the
GBT staff based on measurements of opacity-corrected antenna temperature for
NGC\,7027 at $\sim1.4$\,cm (R. Maddalena, private communication). By
comparing maser line amplitudes among beams and polarizations, we estimate that
the calibration of the $T_{\rm sys}$ is accurate to within $30\%$, an
uncertainty that dominates the error budget for the flux density scale (and is probably due
to calibration uncertainties for the noise sources).  In comparison, antenna
pointing corrections were obtained every $\sim30$ minutes and were typically
$<6''$, which corresponds to a loss less than 8\% in source flux density for a $36''$
beamwidth (FWHM) at 1.3\,cm.  Additional losses due to  pointing errors induced by
varying wind conditions  were less than 20\%, which we estimate using the error model of \citet{Condon2003} 
and the bottom 90\%  of measured wind speeds at GBT weather station 2, for 
each observation.  (Our estimated losses are conservative, considering that 
wind speeds on the feed arm itself were typically lower.)

Data were reduced using custom scripts written in the
Interactive Data Language. We subtracted a running boxcar average of width
$6.25$\,MHz to remove systematic baseline structure from the total-power spectra,
excluding velocity intervals with signal that was detectable above the structure seen in
raw integrations.  The resulting $1\,\sigma$ noise levels attained in an integration of
$\sim30$\,minutes and corrected for atmospheric opacity estimated from
tipping scans (from 0.03 to 0.2) were 4--14\,mJy in a 24.4-kHz
channel. The spectra presented here have been Hanning smoothed (multiple times) 
to a resolution of 108\,kHz, though channel spacing of 24.4\,kHz is shown in the figures.

\newpage
\section{RESULTS}
\label{results}
\subsection{\it Survey---Three New Masers}

In a survey  of 40 inclined ($i>70^{\circ}$) AGN with $v_{\rm sys}<18000$\kms\ selected from the NED (Table\,\ref{survey}), we have detected three new H$_2$O maser sources in Seyfert\,2 systems: NGC\,1320, NGC\,17\footnote{Duplicate designation, NGC\,34, originates from a suspected observer error recorded in \citet{dreyer81} and corrected by \citet{howe00}. Dual designation does not stem from NGC\,17 being a merger with a complicated brightness distribution. However, see discussion of NGC\,17/34 in \citet{hopkins08}.}, and IRAS\,16288+3929 (Figure \ref{new.masers}).   The three discoveries were subsequently confirmed with the Very Large Array (VLA) of the NRAO using a 6.25-MHz observing
bandwidth and 97.7-kHz channels (NGC\,1320, IRAS\,16288+3929) or with the GBT on a
different day but using the same spectrometer configuration (NGC\,17, IRAS\,16288+3929).   Each of the masers is probably associated with nuclear activity.  The NGC\,1320 and IRAS\,16288+3929 maser positions measured with the VLA lie $\ll 1\sigma$ from the optical positions of the nuclei (Table\,\ref{survey}), or less than 300\,pc.   The NGC\,17 maser is three to four times more distant than the farthest H$_2$O maser known to be associated with star formation \citep[Arp\,244;][]{darling08}, yet  it is comparably bright.

Spectroscopic classification of sources as disk maser candidates is most suggestive when three line complexes are detected, one close to the systemic velocity and two bracketing it, with similar but opposite velocity offsets.  Approximate symmetry in this respect is indicative of a disk, and a minimum offset on the order of 200\kms\ is reasonable to exclude misidentification of emission driven by nonnuclear star formation \citep{greenhill2007}.  However, detailed geometry of the underlying accretion disks, orientation with respect to the line of sight, and sensitivity limitations may result in fewer than three detectable line complexes.  In these cases, interpretation of the spectra depends critically on estimates of systemic velocity.

The spectrum of NGC\,1320 is relatively symmetric about a mean of $\sim 2790$\kms\ (Figure \ref{new.masers}), with emission extending outward $\sim 170$\kms\ from peaks at 2614 and 2969\kms.  However, the mean velocity is offset $\sim 70$\,\kms\ from the most precise optical systemic velocity \citep{Huchra1993}, though agreement appears to be good with the estimate of \citet{Bottinelli1992}.  The approximate mirror symmetry and breadth of the line complexes is suggestive of 150--180\kms\ rotation velocity, though it would be twice as great if one of the detected line complexes were to correspond to systemic emission.  Time monitoring might  distinguish between scenarios if a third, blueshifted, variable line complex is detected or if secular drift in velocity is detected in one of the known complexes.

Classification of NGC\,17 is somewhat less certain because of a spread in systemic velocity estimates that spans the separation of the two detected line complexes $\sim 300$\kms\ (Figure \ref{new.masers}).  However, we note that the most precise estimate \citep{rothberg2006}, obtained from dynamical modeling of the galaxy, corresponds well to the peak of the stronger of the two line complexes.  We infer from this circumstance that NGC\,17 is likely to be a disk maser, with rotation velocity $\sim 300$\kms.    As for NGC\,1320, monitoring may increase certainty should a blueshifted line complex  or secular drift be detected.   Nonetheless, because NGC\,17 is a merger \citep[e.g.,][]{Vorontsov1977,Mazzarella1993,Hunt1999,Schweizer2007}, interpretation of the spectrum requires some caution.  The velocity offset between the two spectral features is comparable to the stellar velocity dispersion, $\sigma_\circ=201\pm8$\kms\ \citep{rothberg2006}, and the maser lines may arise in different compact structures within the system.

For IRAS\,16288+3929, classification is problematic because only a single spectral feature is evident ($\sim8920$\kms), and it is displaced $\sim200$\kms\ from available optical systemic velocities (Figure \ref{new.masers}).  The emission may represent high-velocity blueshifted emission or activity unrelated to any accretion disk.  An alternative origin could be jet activity driven by the central engine. We note comparable velocity offsets for known ``jet masers'' \citep[NGC\,1068, NGC\,1052, Mrk\,348, and M\,51; ][]{Gallimore1996, Gallimore2001, Claussen1998, Kameno2005, Peck2003, Hagiwara2007}.  On the other hand, in these cases, spectral features are characteristically at least 50\,km\,s$^{-1}$ wide, at least an order of magnitude broader than the  linewidth for the known maser feature in IRAS\,16288+3929.

NGC\,1320 and NGC\,17 were targeted but not detected in previous surveys for
maser emission.  Braatz, Wilson, \& Henkel (1996)\nocite{Braatz1996} and Kondratko et al. (2006a)\nocite{Kondratko2006a} report observation of NGC\,17 with 
$1\sigma$ noise levels of  73  and 16\,mJy in  0.66  and 1.3\kms\ spectral channels, respectively.  These correspond to signal-to-noise ratios (SNR) of $\la 1$ for the observed line flux
of $11$\,mJy (peak) over 1.5\kms.  In the case of NGC\,1320,
these earlier studies obtained $1\sigma$ noise levels of $61$ and $19$\,mJy in $0.84$ and $1.3$\,km\,s$^{-1}$ spectral channels, respectively.  The SNR for the observed line flux of $37$\,mJy, at $1.5$\,km\,s$^{-1}$ resolution, would have been less than two.

\subsection{\it Three New Disk Maser Candidates (NGC\,449, NGC\,2979, and NGC\,3735) among Known Sources}

Through deep integration, we detected new features in the spectra of masers not previously recognized to have high-velocity emission and evidence for origins in accretion disks (Figures\,\ref{other1}, \ref{other2}, \ref{other3}; Table\,\ref{TableOther}).  Noise levels were 2--4\,mJy ($1\sigma$) referenced to outside the atmosphere.  We comment on individual systems below. The first three we discuss are the new disk maser candidates.

{\it NGC\,449 (Mrk\,1)} (Figure \ref{other1}). We detected an emission complex at
$\sim4860$\,km\,s$^{-1}$, a $\sim20$\,mJy component at $\sim4600$\,km\,s$^{-1}$,
and a new weak line complex (7\,mJy peak in a $1.5$\,km\,s$^{-1}$ spectral channel) at
$\sim4720$\,km\,s$^{-1}$. The complex at $\sim4860$\,km\,s$^{-1}$ has been
present in the spectrum of the source since its discovery
\citep{Braatz1996,Braatz2003} while the component at $\sim4600$\,km\,s$^{-1}$ was
marginally detected first by \cite{Braatz2003}. The newly detected emission is
approximately halfway in velocity between the two previously known spectral
complexes. Although the position of the maser emission with respect to the
nucleus has not been measured, the spectrum shows the archetypical spectral
signature of emission from a highly inclined accretion disk. In particular, considering
the uncertainty on the systemic velocity of the galaxy (Figure\,\ref{other1}; Table\,\ref{TableOther}), we associate the complex near 4720\,\kms\ with low-velocity disk emission and infer an orbital velocity (which we take to be the maximum velocity with respect to the systemic velocity) of $\sim150$\,\kms. It is notable that the suggested systemic velocity of the galaxy would be blueshifted by a few times 10\,\kms\ from optical measures and $\sim 100$\,\kms\ from the estimate obtained from HI data.

{\it NGC\,2979} (Figure \ref{other1}). Emission was newly detected at a level of $\sim3$\,mJy at
2390\,\kms\ \citep[cf.][]{Greenhill2003survey}, which is blueshifted by
$\sim300$\,\kms\ from estimates of the  systemic velocity.  The new spectrum is approximately centered on the optical systemic velocity estimate of $2720\pm15$\,\kms\ \citep{Fisher1995}, which is suggestive of emission from a highly inclined disk with $\sim 300$\,\kms\ orbital velocity, though extended line wings may suggest a somewhat more rapid rotation.
This interpretation is consistent with the known coincidence of the maser source position and the active nucleus \citep{Greenhill2003survey}.  Asymmetry in the maser spectrum is pronounced but difficult to explain without imaging data that would enable geometric  modeling of the disk.  However, we note that the archetypal disk maser in NGC\,4258 exhibits about a 10 times imbalance in the flux densities of red- and blueshifted emission.   In the new spectrum, the velocity range of emission $\ga 10$\,mJy is consistent with a weak plateau observed in the 6 times lower sensitivity discovery spectrum \citep{Greenhill2003survey}.

{\it NGC\,3735} (Figure \ref{other1}). We detected narrow lines at  2444, 2484,
2784, and 2896\,\kms\ that were  not observed in previous spectra \citep{Greenhill1997b,Braatz2003}.  
An additional complex of weak lines may be present between 2830 and 2870\,\kms.  Doppler shifts from systemic velocities derived from HI data are  $\pm$ 200--250\,\kms\ \citep{Staveley1988}.  
This bracketing of systemic emission and known angular coincidence of the maser and nucleus 
\citep{Greenhill1997b} are suggestive of emission from a close-to edge-on accretion disk.

{\it UGC\,4203 (Mrk\,1210)} (Figure \ref{other2}). We detected line complexes bracketing the systemic velocity, at 3755--3861 km~s$^{-1}$ and 4190--4350\,\kms\ (Figure\,\ref{other2}), as well as isolated narrow lines distributed between these, at 4106\,\kms\ ($\sim 70$\,mJy),  4134\,\kms and 3992\,\kms\ (5--10\,mJy).  The peaks of the red- and blueshifted complexes were  detected with lower signal-to-noise ratio previously \citep{Braatz1996, Braatz2003}, though the symmetry of the spectrum around the systemic velocity (Table\,\ref{TableOther}) and suggestion of emission from an inclined disk were less pronounced.   The inferred orbital velocity is $\sim 300$\,\kms, depending on the adopted systemic velocity.  

{\it Arp\,299 (NGC\,3690)} (Figure \ref{other2}). The spectrum contains a broad feature with 
FWHM\,$\sim 230$\,km\,s$^{-1}$ and overlying narrow Doppler components that were not evident in the previous spectra of the source obtained with $4.3$\,\kms\ channels 
\citep{Henkel2005}. Maser emission is known to arise from three regions in this merging system, the nuclei of NGC\,3690 and IC\,694 and an internuclear radio continuum ``hotspot''  \citep{tarchi07}.  Both nuclear maser sources are $\sim200$\,\kms\ wide, which is notable because large widths have been associated at least circumstantially with jet activity \citep[see][]{tarchi07}, as exemplified in three AGN where maser and radio continuum emission have been mapped relative to one another interferometrically: NGC\,1052, NGC\,1068, Mrk\,348 \citep[][also review by Greenhill 2007]{Claussen1998, Gallimore2001, Peck2003, Kameno2005}.  Notable time variability has been observed toward these masers, and position-resolved monitoring of Arp\,299 may offer useful diagnostics.

{\it NGC\,4293} (Figure \ref{other2}). The spectrum exhibits emission distributed over $\sim 290$\,\kms\ but that unevenly brackets the most precise estimate of systemic velocity \citep{diNella1995}.  An additional emission feature at $\sim 1200$\kms\ may be present at the $\sim 3\sigma$ level.
The case for origin in a disk is marginal due to uncertainty in the full velocity range of emission and the degree of symmetry.

{\it NGC\,0235A} (Figure \ref{other3}). Dominated by a single spectral feature redshifted by $\sim 300$\,\kms\ from the systemic velocity, interpretation of the spectrum is difficult.  Goals of further monitoring should  include a search for blueshifted emission.  

{\it NGC\,2824} (Figure \ref{other3}). As observed in the case of Arp\,299, the NGC\,2824 maser exhibits narrow Doppler components overlying a broad component (FWHM $\sim150$\,km\,s$^{-1}$). This is 
consistent with the discovery spectrum \citep{Greenhill2003survey}, though 
the newly recognized zero-intensity width is substantially broadened, $\sim 450$\,\kms.  
The maser emission is associated in position with the AGN
\citep{Greenhill2003survey}, and the spectrum is suggestive of jet activity, but the interpretation is uncertain.

{\it IRASF\,19370-0131} (Figure \ref{other3}). Observed emission is limited to a range $\sim 50$\,\kms\ centered on the systemic velocity estimated by \citet{Strauss1992}.  Detection of secular drift in feature velocities,  as well as emission over a broader range would be useful goals for future monitoring.

\subsection{\it New Centripetal Accelerations for Three Disks}

Deep integrations were also used in monitoring three masers reported by Kondratko, Greenhill, \& Moran (2006b)\nocite{Kondratko2006b}, which display among the largest Doppler shifts for high-velocity emission: CG\,211, NGC\,6264, and
VV\,340A  (Table\,\ref{accel.log}; Figure\,\ref{cumulative.3panel}). We detected secular drifts in low-velocity features, which probably reflect centripetal acceleration due to disk rotation (Table
\ref{gauss_fit_table}, Figs.\,\ref{gauss_fit1} and \ref{gauss_fit2}). For each source, we used an iterative least-squares technique, adopted from \citet{Humphreys2008}, to decompose the time-series spectra into Gaussian components, solving simultaneously for the light curve, time-varying line width, velocity at a reference epoch, and velocity drift  of each component.  (Each component was assumed to have its own fixed drift rate.)  We constrained the widths of most Gaussian components to be $<5$\,\kms, motivated by empirical observation.  A first-order polynomial and a wide ($\sim25$\,km\,s$^{-1}$) stationary Gaussian component were used to model the broad low-velocity plateaus in VV\,340A and CG\,211, respectively. The fitted accelerations were not sensitive to inclusion of the broad
components, though reduced $\chi^2$ estimates were improved.  We obtained drifts of  0.1--1.9\,\kms\,yr$^{-1}$ with uncertainties of 0.005--0.09\,\kms\,yr$^{-1}$, 1.2--2.1\,\kms\,yr$^{-1}$ with uncertainties of 0.1--0.6\,\kms\,yr$^{-1}$, and $3.8\pm 0.1$\,\kms\,yr$^{-1}$ for CG\,211, NGC\,6264, and VV\,340A, respectively (Table \ref{gauss_fit_table}).

\section{DISCUSSION}
\label{discussion}

\subsection{\it Black Hole Masses, Disk Radii, and Their Implications}

Mean measured centripetal acceleration, $ a $, combined with mean orbital velocity inferred from single-dish spectra, $ v $, can be used to estimate central-engine gravitational masses and accretion disk radii. The mean acceleration among low-velocity features may be expressed as $ a_l = v_l^2/R_l$, where $v_l$ is an effective mean rotational velocity and $R_l$ is the effective mean radius, both for material contributing low-velocity emission.  The mean rotational velocity inferred from spectra may be expressed as $ v_h^2=GM/R_h$, where $G$ is the gravitational constant, $M$ is black hole mass, and $R_h$ is the effect mean radius for material contributing high-velocity emission.  For edge-on orientation, we obtain $M= v_h ^2 v_l^2 (R_h/R_l)/G a_l $. Unfortunately, without VLBI, we cannot determine $R_h$ and $R_l$ independently or, therefore, $v_h$ and $v_l$ independently. Hence, for the purpose of calculation, we assume that the low- and high-velocity emission arise at the same effective mean radii ($R_h\sim R_l\equiv R$) and that the effective mean rotation velocity of accelerating low-velocity material is the same as the measured mean rotational velocity, $v_h \sim v_l\equiv v$. We then obtain $M = v ^4 / G a_l $ and $R=v^2/a$. These values are reported in Table~\ref{mass.radius}. VV\,340A may have the largest central mass among known maser galaxies.

We note that the scatter in the velocity drift measurements for CG\,211 is substantially larger than the individual formal uncertainties. Since $a=GM/r^2$ for low-velocity material and hence $\Delta a/ a =-2\,\Delta r/ r $, the scatter in the measured accelerations of $\Delta a/ a\sim 1.0$ might be due to low-velocity emission that populates a wide range of radii within the accretion disk, i.e., $\Delta r/ r \sim 0.5$. In our simple model, the radial spread among high-velocity material might be expected to be $\Delta v/ v =0.5\,\Delta r/ r \sim0.25$. The observed value of $\Delta v/v$ in the high-velocity components is about 0.3 (see Figure 5). This suggests that the radial spread is about the same for both high- and low-velocity masers.  For NGC\,6264, $\Delta a/a\sim0.4$, so $\Delta r/r\sim0.2$, and $\Delta v/v$ for the high-velocity masers is expected to be about 0.1. However, inspection of the spectrum (Figure \ref{cumulative.3panel}) suggests that $\Delta v/v\sim0.5$. Hence, the high-velocity masers probably occupy a larger range of radii than the low-velocity masers. Note that the case of NGC\,6264 may be similar to that of NGC\,4258, where the radial range of the high-velocity features is also greater than that of the low-velocity features \citep{Herrnstein2005}

In each case where we report centripetal acceleration, it has been positive, i.e., spectral components drift towards larger velocities.  This is consistent with the  emission being from material on the near side of the disks.  Material behind a dynamical center, which would exhibit negative centripetal accelerations, has not been detected toward any known disk maser or disk maser candidate. Although the relatively low signal-to-noise achieved for CG\,211, NGC\,6264, and VV\,340A would make recognition of  ``back-side'' emission relatively difficult, we can exclude the possibility that flux densities are equivalent to the front-side emission. It is difficult to know whether the apparent absence of back-side emission  may be due to a maser radiation propagation effect (e.g., competitive gain diminishes inward-directed beams) or differential  absorption  (e.g., inward-directed beams pass through regions of enhanced scattering or absorption in the vicinity of a central engine; inward-directed beams cross the full breadth of a disk and are attenuated by high emission-measure material at large disk scale heights, as in Herrnstein, Greenhill, \& Moran 1996\nocite{Herrnstein1996}).

The three masers in Table\,\ref{mass.radius} are priority targets 
for interferometric programs that are intended to estimate geometric distances (e.g., Herrnstein et al. 1999) and (in combination with distances for other maser targets) a combined Hubble constant, H$_0$.
For a sample of $N$ anchor galaxies, uncertainty in H$_0$  scales as $\sim N^{-0.5}$ (for uniform uncertainties) because errors for individual distance measurements are largely uncorrelated; they depend principally on the distribution of Doppler components across the disk face and details of disk geometries, which differ from galaxy to galaxy.  

To provide substantive
independent constraint on additional cosmological parameters, such as a time-variable equation of 
state (EOS) for dark energy and curvature, estimates of H$_0$ need to achieve on the order of 1\% uncertainty, matching the growing accuracy of other measurements, e.g., Cosmic Microwave Background (CMB) fluctuations and Baryon Acoustic Oscillations \citep[][and references therein]{hu05,it08,Olling2007,greenhill09}. In combination with CMB data alone, a 1\% estimate of H$_0$ would enable a 5\% constraint on a presumed-constant EOS, for a flat universe \citep[Figure 14,][]{riess09}. 
A sample on the order of 100 disk masers would be required to achieve 1\% accuracy in H$_0$,
if each distance is accurate to 10\%. At present, this is at least an order of magnitude beyond the number of disk masers apparently 
suitable for distance measurement.  However, a three-year survey of 3000 galaxies at the GBT, achieving RMS noise $\sim 3$ mJy over 1\,\kms\  (4 sources per hour for $\sim 750^h$), would result in the detection of 150 masers, assuming a conservative detection rate of one-half that reported by 
\citet{Kondratko2006b} for AGN  at 10000--15000\,\kms.  For a disk maser fraction of one-third, and a further one-quarter that appear suitable for distance measurement (as suggested in the currently known population), it would be possible to assemble a sample on the order of 25 objects. For 
10\% distance uncertainty per maser, this would enable estimation of H$_0$ with 2\% accuracy in the near term.   Prospects are improved by better individual distances as may be achievable with the VSOP-2 space-VLBI mission  and larger target samples made accessible by the Square Kilometer Array \citep{greenhill04, morganti04}.

However, peculiar radial motions of galaxies increase scatter in the Hubble relation and introduce partial correlation among recessional velocities. The estimate of the Hubble constant from any particular galaxy measurement is $D/v$. The sample size needed to achieve a given total uncertainty grows 
as the quadrature sum of  $\Delta D/D$ and fractional peculiar motion, $v_{pec}/v_{sys}$, so that the uncertainty in a sample of $N$ galaxies is 
$\Delta H_0/H_0 \sim {[{(v_{pec}/v_{sys})}^2 + {(\Delta D/D)}^2]}^{1/2}~N^{-1/2}$. 
If $\Delta D/D=0.1$, and we require that the uncertainty in peculiar motion raises the error by less than 
10\%, then $v_{pec}/v_{sys}<0.05$. Outside of clusters, $v_{pec} \la 500$\kms\ is typical \citep{springob07}. We take this to be an RMS value and infer that $v_{sys} \ga10^4$\,\kms~is needed.  For maser hosts inside rich clusters, barycenter velocities can be  estimated with uncertainty $<200$\,\kms\ \citep[e.g.,][]{crook07} and  used in place of individual galaxy recessional velocities, though cluster peculiar motions would still deserve attention.

It would be desirable to target relatively nearby galaxies that are in ``quiet'' portions of the Hubble flow, 
to reduce statistical and nonstatistical errors \citep[e.g., UGC\,3789;][]{masters05,reid09}. 
However,  a broad distribution of sample members in velocity and angle on the sky is needed.  
Subtraction of cosmic flow models to at least partially correct for $v_{pec}$ reduces the 
lower limit on acceptable $v_{sys}$ and admits galaxies in not so quiet regions.   Using Tully-Fisher distances to estimate peculiar velocities and construct a multi-attractor model inside 6000\kms, 
\citet{masters05} obtained a $160\pm20$\,\kms\ RMS residual. \citet{Erdogdu2006}  
report a somewhat less direct error estimate, a scatter of 100--150\,\kms\ around a flow model 
derived from galaxy recessional velocities (as opposed to peculiar velocities), up to $\sim 16000$\,\kms.
Adopting a characteristic 160\,\kms\ uncertainty, the lower limit in $v_{sys}$ is  3200\,\kms,
for maser distances good to 10\% and the nominal limit of $\sim 10\%$ contribution by peculiar motions to the error budget.
The technique works for maser hosts away from the Great Attractor and the galactic plane ($|b|>20^\circ$; Masters 2005\nocite{masters05}), where flow models are not well constrained.

\subsection{\it Accretion Disk and Host Galaxy Misalignments}

In studies of the relative orientation of AGN central engines and host galaxies, radio jet position angles at parsec and kiloparsec scales are used to infer the orientations of the accretion disks that launch them.  Ionization cone position angles are used to infer the orientations of larger scale, dusty, geometrically thick structures at  larger  radii.  Comparisons of position angles suggest that accretion disks are not aligned with galactic spiral disks \citep{Ulvestad1984, Nagar1999, Kinney2000, Schmitt2003, Middelberg2004}.  At the same time, the accretion disks appear reasonably well aligned with the structures that collimate ionization cones \citep{Schmitt2003}, though the sample is small.

Because interferometric mapping has demonstrated that disk maser emission marks molecular structures that are oriented close to edge-on, the emission provides a relatively precise indicator of orientation, via inclination rather than position angle, as is otherwise the case principally for radio jets and ionization cones. Thus, estimation of stellar disk inclination in spiral galaxies that host disk masers enables immediate estimation of (mis)alignment  between parsec-scale and kiloparsec-scale structures.  The approach complements other techniques, in terms of orientation angle (position angle vs inclination) and of radius, where the range of radii for maser emission lies outside the zone where accretion disks act to collimate radio jets. Perhaps the fullest case studies would be possible for galaxies with observable radio jets, disk maser emission, well-resolved ionization cones, accurate B-band isophotal radii (for  detection of spiral arms), and a reasonably firm morphological type.  Among nearby galaxies, inclinations could be obtained from detailed modeling of spectroscopic data for well-resolved stellar disks and  circumnuclear molecular rings \citep[e.g.,][]{Curran2000}.

A suitable sample for study may be compiled directly from lists of known disk maser candidate hosts that are spirals. The spectroscopic identification of masers that trace underlying disk structure is sufficient. Galactic inclinations, to first order, may be inferred directly from axial ratios, $i = \cos^{-1}(b/a)$, though detailed dynamical modeling would be preferred to obtain the most accurate values.
  
Assuming the spectroscopic identifications are accurate, then there is no apparent preponderance of edge-on galactic disks among disk maser candidate hosts. For  inclinations $< $70--80$^\circ$,  the distribution of  orientations for the disk maser candidate matches what would be anticipated for random galaxy orientations (Figure~\ref{incl.histo}).  To make the comparison, we assembled a sample of axial ratios for published disk maser candidates, defined with respect 
to surface brightness of $B=25.0$ per sec$^{-2}$, for S0 and later systems (Table~\ref{tab.incl}).  Axial ratios were taken from the Third Reference Catalogue of Bright Galaxies  \citep{Vaucouleurs1991}.  

The apparent undercount among high-inclination systems in Figure\,\ref{incl.histo} is probably due to overestimation of disk minor axes due to galactic bulges, which is anticipated to be greatest for nearly edge-on massive systems (i.e., morphologies from S0 to Sa).    
\citet{fouque90} describe an empirical correction that is dependent on galaxy morphology, quantified in the $T$ parameter \citep[e.g.,][]{Vaucouleurs1991}.  Determination of morphologies is strongly dependent on image quality and may be affected by subjective bias \citep[e.g., see methodology discussion of][]{fukugita07}.  Corrected inclinations are particularly sensitive to uncertainty in $T$  for highly inclined systems, where forbidden values of $\cos i$ may result.  We estimate corrected inclinations using catalogued morphological types \citep{Vaucouleurs1991, Malkan1998, lauberts89}.  For galaxies with forbidden values of $\cos i$ (NGC\,1386, NGC\,4388, NGC\,3735), we increment $T$ by 0.6, which is approximately the catalogued uncertainty for these systems.  We obtain inclinations that are within a few degrees of $90^\circ$. (In light of the chosen coarse binning, the precise values are not critical.)   As a result, agreement with what is expected for random galactic disk orientations is obtained over the whole range  (Figure\,\ref{incl.histo}). For an intensively studied sample that includes well-characterized inclination uncertainties for each object, formal statistical comparison would be possible using a Monte Carlo analysis.

No correlation of misalignment with morphological type or mass is apparent.  Disk maser candidate hosts span a range of almost two orders of magnitude in black hole mass and a broad range of apparent morphological type, from S0 ($T=-2$) to Sc ($T\le5$), excluding the merger system NGC\,17.  The median type is 2.7, with $1<T<4$ for the middle quartiles  (Table~\ref{tab.incl}).  Masses are inferred from VLBI or acceleration measurements: NGC\,4258 \citep[e.g.,][]{Miyoshi1995, Herrnstein2005}, NGC\,1068 \citep{Greenhill1997}, Circinus (Greenhill et al. 2003a), NGC\,3079 \citep{Yamauchi2004, Kondratko2005},  UGC\,3789 \citep{reid09}, NGC\,6323 \citep{Braatz2007},  IC\,2560 \citep{Ishihara2001}, Mrk\,1419 \citep{henkel02}, and CG\,211, NGC\,6264, and VV\,340A (this work).  

Comparison of galactic plane misalignments inferred for disk masers and for accretion disks at smaller radii thus far is statistical.  Direct comparison would require disk masers that have been mapped with VLBI, but that sample is comparatively small.  In several examples, the position angles for jets (or lobes) and ionization cones are well aligned with disk maser rotation axes,  projected on the sky:  NGC\,3393 (Kondratko, Greenhill, \& Moran 2008)\nocite{kondratko08}, NGC\,4258 \citep{Cecil2000}, and NGC\,4945 \citep{Greenhill1997c}.  However, misalignment between a jet and disk maser has also been observed in the case of NGC\,3079 \citep{Kondratko2005}, and skewing of ionization cone orientation by shadowing from a warped disk has been observed  in Circinus \citep{Greenhill2003}, both at radii $\la 1$\,pc. Misalignment on the order of $30^\circ$ is also observed in NGC\,1068, between the position angle of the disk maser and the major axis of ionized  material just inside the inner radius for molecular emission  \citep{Gallimore2004}.  If these examples are representative, counterparts may contribute to scatter in the alignment of jets and ionization cones noted by \citet{Schmitt2003} but ascribed to signal-to-noise and projection effects.  
Warping as observed among disk masers and larger nuclear structures \citep[e.g.,][]{schinnerer00} may also be responsible, where fueling of central engines does not depend on conveyance of material along galactic planes to ever smaller radii and vertical heights until it is concentrated into a coplanar accretion disk.

\section{CONCLUSIONS}
\label{conclusions}

\begin{figure}[p]
\epsscale{0.75} \plotone{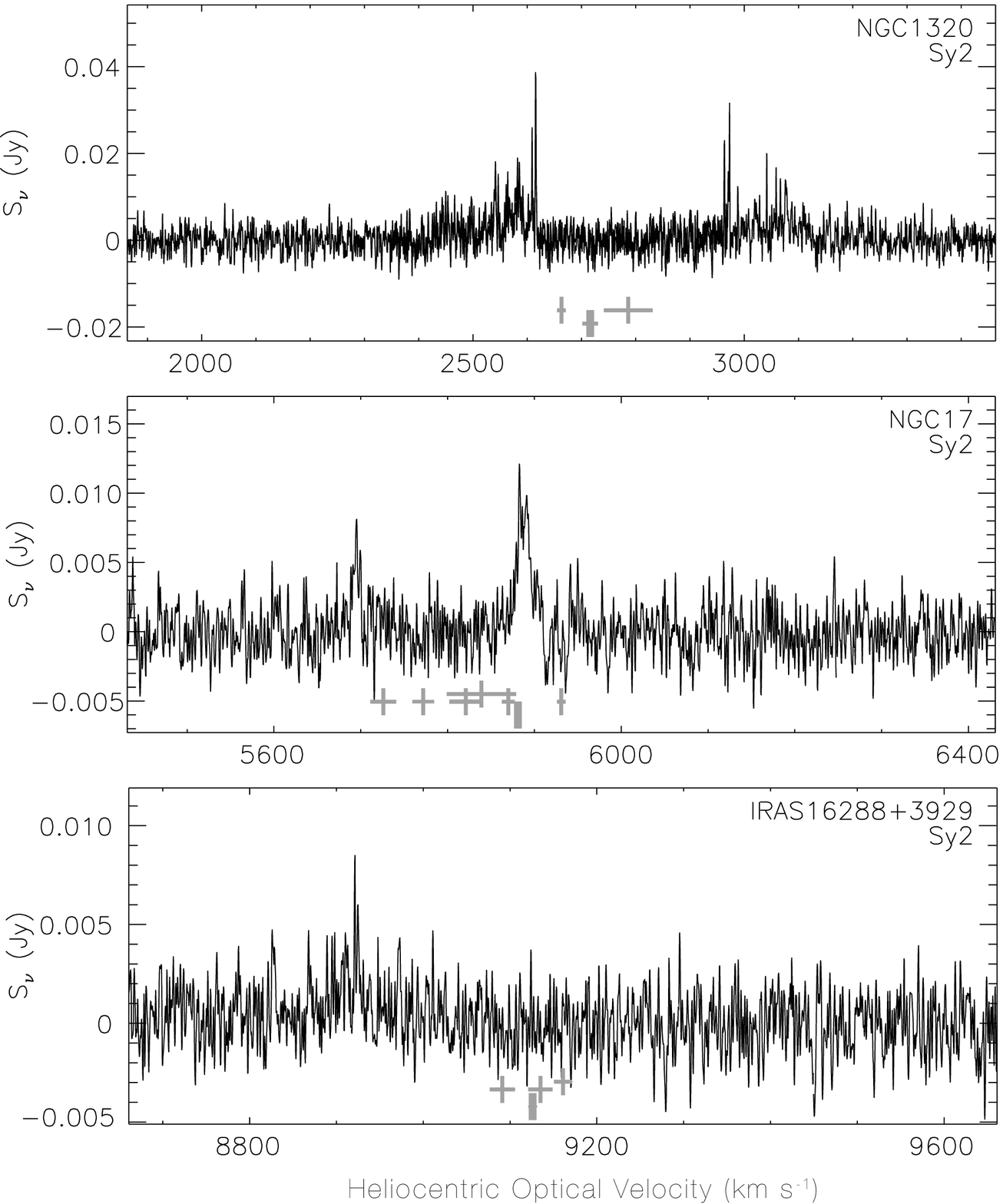}
\caption{Spectra of newly discovered emission in NGC\,1320, NGC\,17, and
IRAS\,16288+3929.  After smoothing, the spectral resolution 
is 108 kHz (1.5\,\kms), for which $1\sigma$ noise levels are 2.8, 1.5, and 1.3\,mJy, respectively.
Vertical bars indicate systemic velocity estimates. Where there is more than one, the thicker one represents the adopted value, selected based on consideration of error budget.  Velocity estimates for NGC\,1320 are $2716\pm 29$ \citep[optical;][]{Huchra1993}, which is adopted here, $2663\pm 16$ \citep[HI;][]{Davoust2004}, and $2786\pm90$ \citep[optical;][]{Bottinelli1992}.  (We note duplication of NGC designation, and the swapping of MCG identifier and velocity between NGC\,1320 and NGC\,1321 entries in da Costa et  al. 1998.)  For NGC\,17, the velocities are $5881\pm2$ \citep[optical;][]{rothberg2006}, which is  adopted here, $5931\pm10$ \citep[HI;][]{Bottinelli1990}, $5772\pm 25$ \citep[optical;][]{Vaucouleurs1991}, $5726\pm30$ \citep[optical;][]{Osterbrock1983}, $5821\pm 38$ \citep[optical;][]{Huchra1993}, $5821\pm 44$ (optical; da Costa et al. 1998), $5870\pm15$ \citep[optical;][]{Schweizer2007}, and $5839\pm80$ \citep[optical;][]{Jones2005}.   For IRAS\,16288+3929, optical velocities are $9123\pm9$ (emission lines; adopted here) and $9136\pm27$ (absorption lines), both from \citet{sdss3}, and $9091\pm29$ \citep[][]{Zabludoff1990} and  $9161\pm21$ \cite[][]{Rines2002}.}

\label{new.masers}
\nocite{daCosta1998}
\end{figure}

\begin{figure}[p]
\plotone{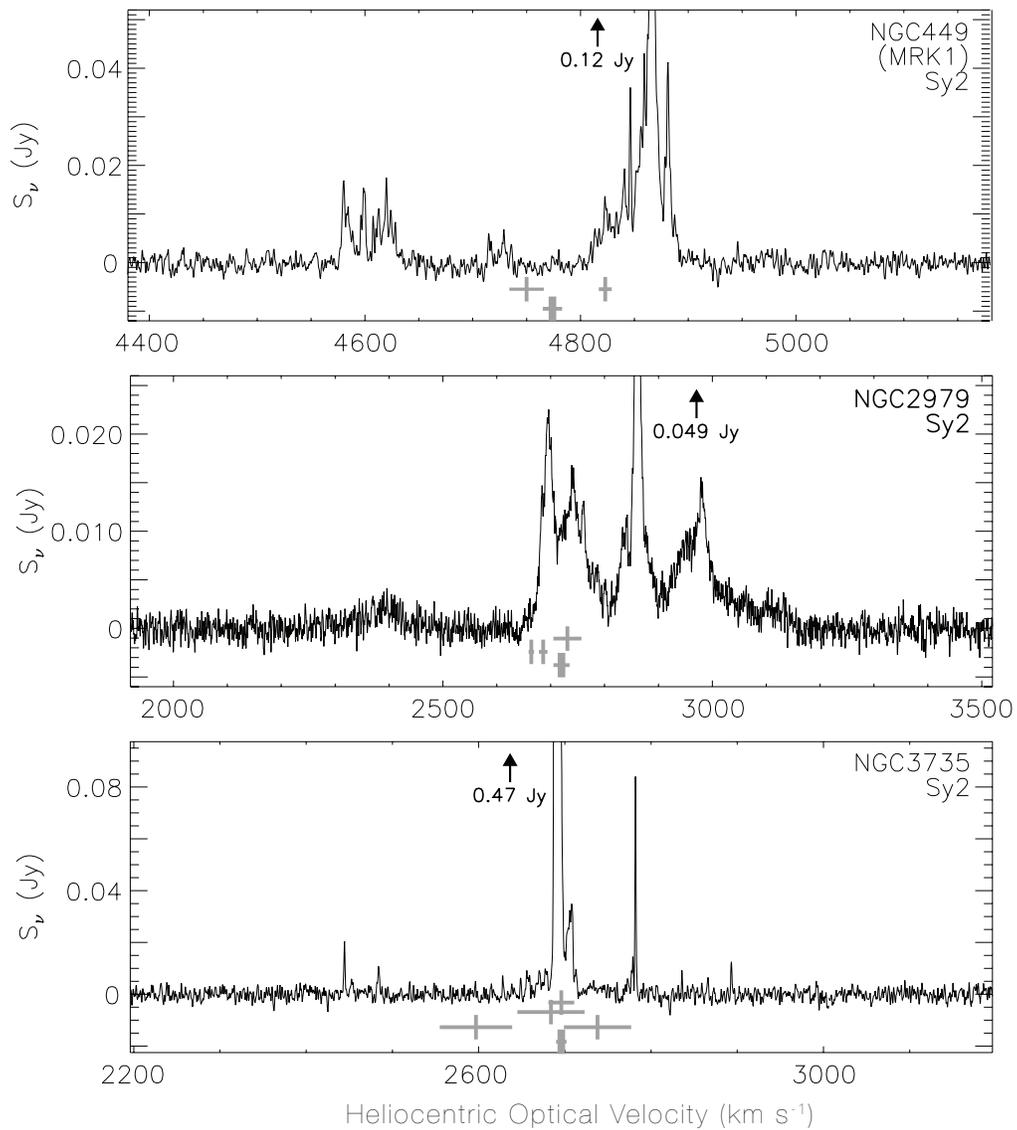} \caption{Spectra of known masers in NGC\,449 (Mrk\,1), NGC\,2979, and NGC\,3735, obtained with the GBT.  After iterative Hanning smoothing, the effective spectral resolution is 1.5\,\kms.   Vertical bars indicate systemic velocity estimates; where there are more than one, the thicker one represents the adopted value (see Table~\ref{TableOther}).  The vertical scale was expanded to make apparent weak emission.  The peak flux densities are noted below the vertical arrows.} 
\label{other1}
\end{figure}

\begin{figure}[p]
\plotone{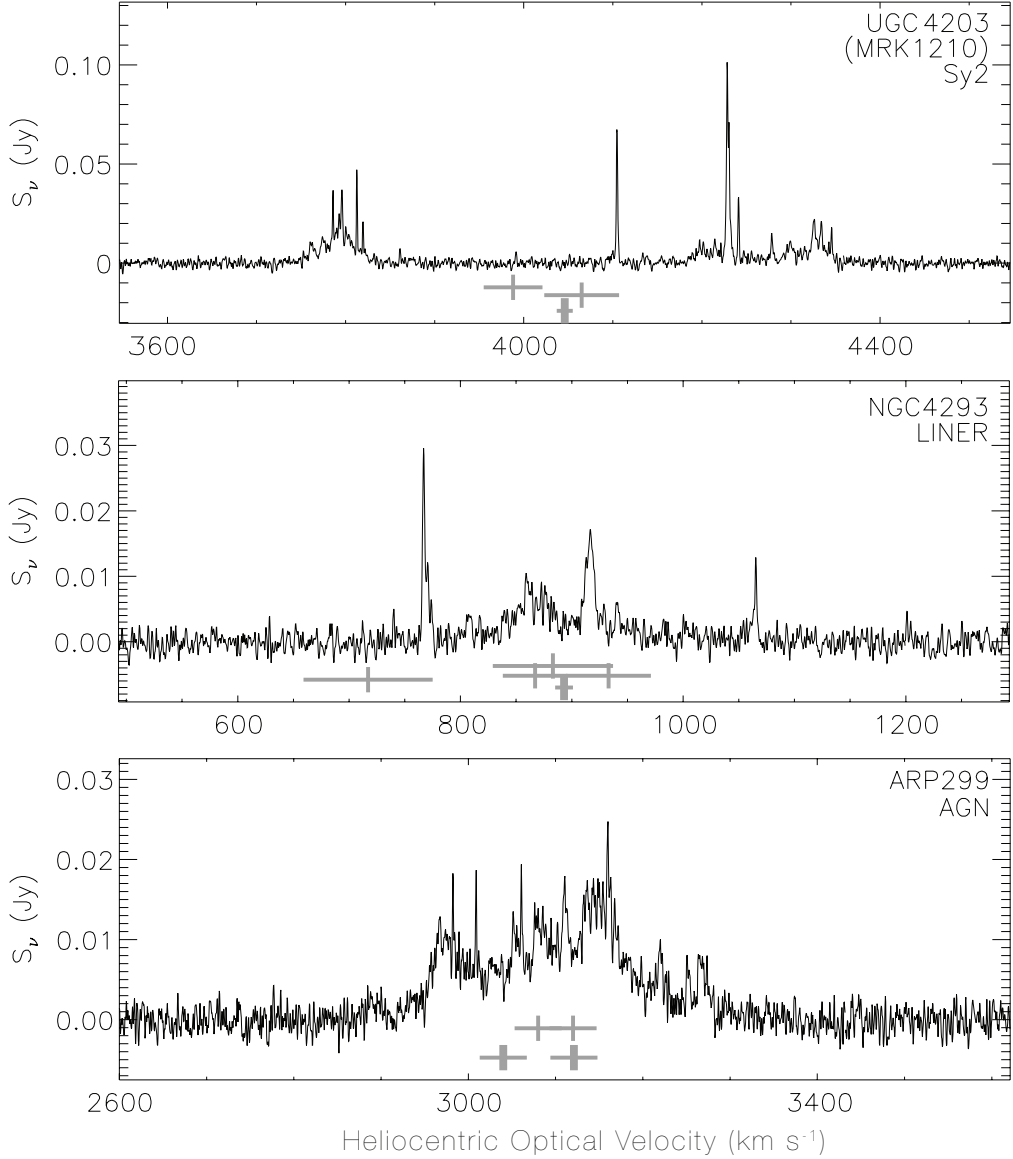} 
\caption{Spectra of known masers in UGC\,4203, Arp\,299, and NGC\,4293, as in Figure~\ref{other1}.  For Arp\,299, systemic velocities are shown for each nucleus, NGC\,3690 and IC\,694.  See \citet{Hibbard1999} for discussion of nomenclature and identification.}
\label{other2}
\end{figure}

\begin{figure*}[p]
\plotone{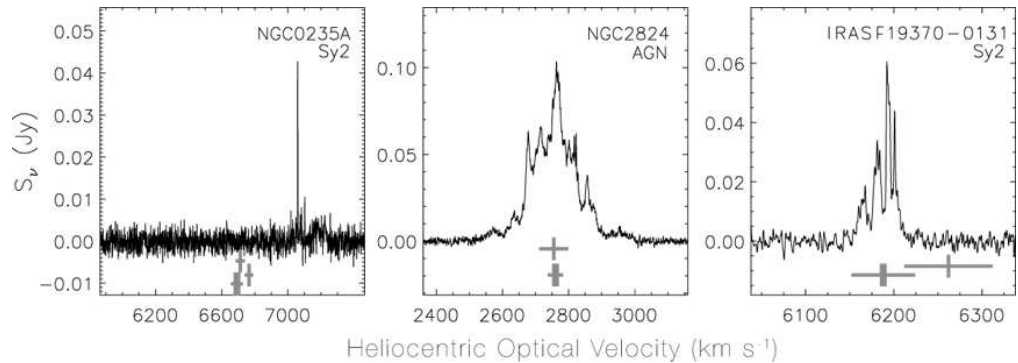} 
\caption{Spectra of known masers in NGC\,0235A (ESO474-G016), NGC\,2824, and 
IRASF\,19370-0131, as in Figure\,\ref{other1}.  NGC\,0235A is part of a close pair ($\sim 11''$) 
with ESO474-G017, identified as a Seyfert\,2 object in the NED.  The systemic velocity $6772\pm86$\,km\,s$^{-1}$ (da Costa et al. 1998) is also close to the maser line velocity. }
\nocite{daCosta1998}
\label{other3}
\end{figure*}

\begin{figure}[p]
\plotone{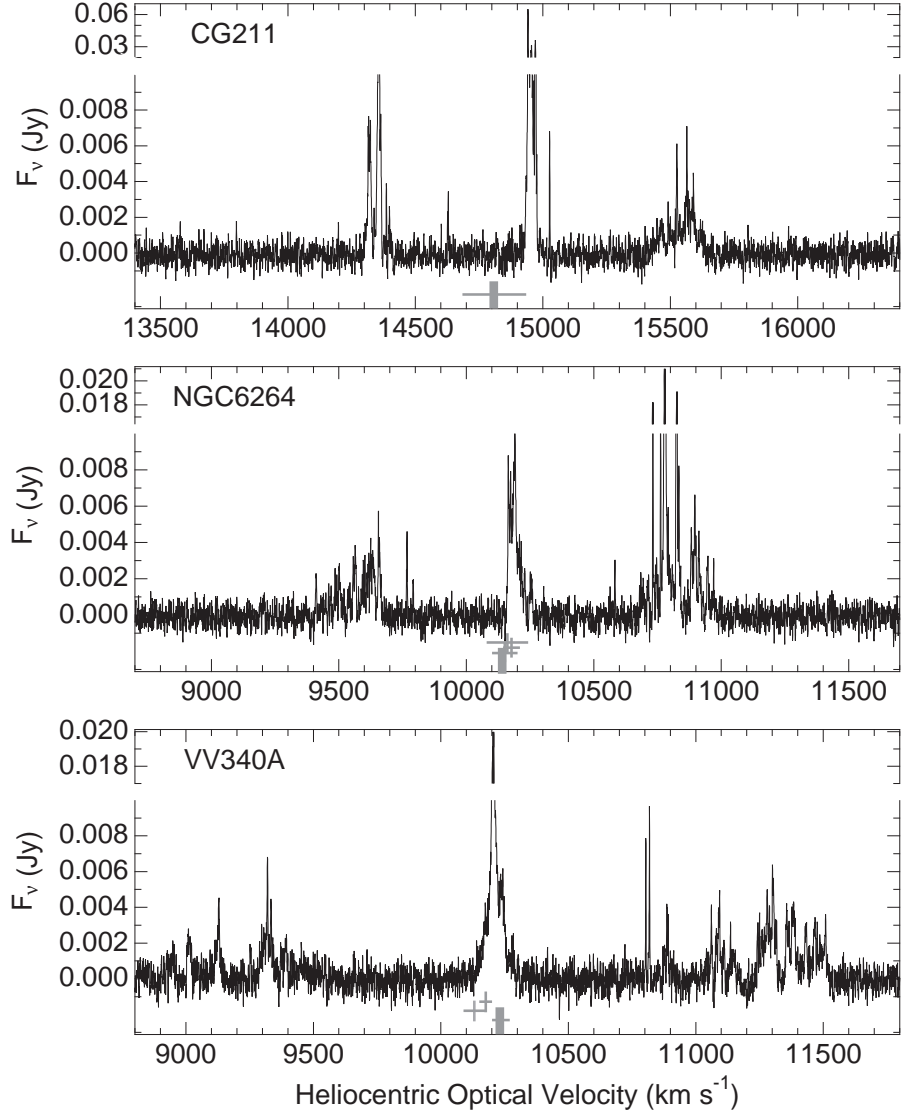}
\caption{Variance-weighted average spectra for CG\,211, NGC\,6264, VV\,340A, for the epochs listed in Table\,\ref{accel.log}. Spectra have been iteratively Hanning smoothed to a resolution of 1.5\,\kms, resulting in 1-$\sigma$ noise levels of 0.53, 0.42, and 0.63 mJy, respectively. The vertical axes are split to emphasize Doppler components below 10\,mJy.  Vertical bars indicate systemic velocity estimates; where there is more than one, the thicker one represents the adopted value. CG\,211: $14810\pm120$\kms\ \citep[optical;][]{weistrop1991}; NGC\,6264: $10141\pm10$\,\kms\ \citep[optical;][]{sdss5}; $10177\pm28$\kms\ \citep[optical;][]{beers1995},  $10161\pm76$\kms\ \citep[optical;][]{koranyi2002}, $10151\pm45$\kms\ \citep[optical;][]{falco1999}; VV\,340A:  $10234\pm29$\kms\ \citep[optical;][]{falco1999}, $10131\pm75$\kms\ \citep[optical, mean and range;][]{keel96}, $10176\pm20$\kms\ \citep[optical;][]{kara1980}.}
\label{cumulative.3panel}
\end{figure}

\begin{figure}[p]
\epsscale{1.1}
\plottwo{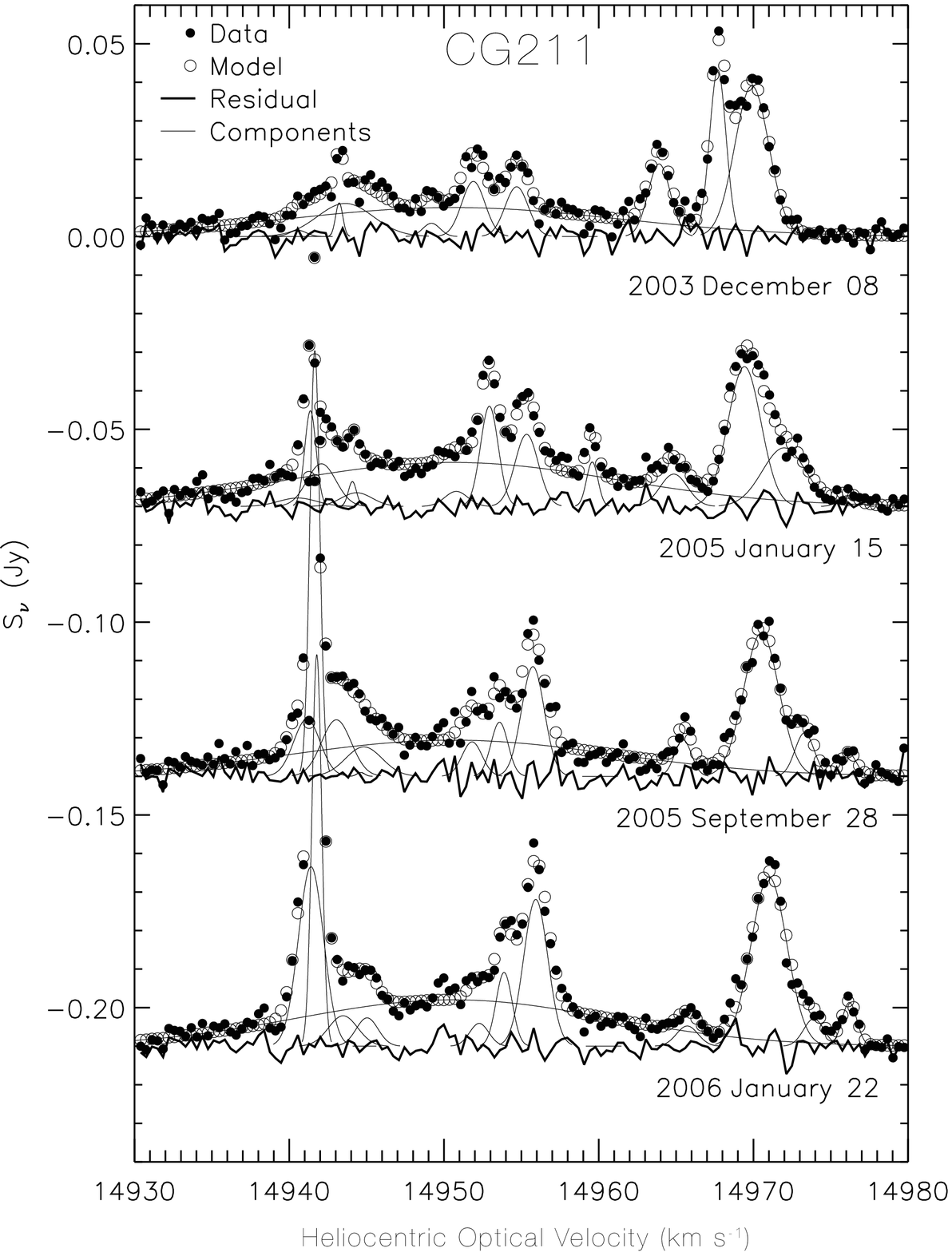}{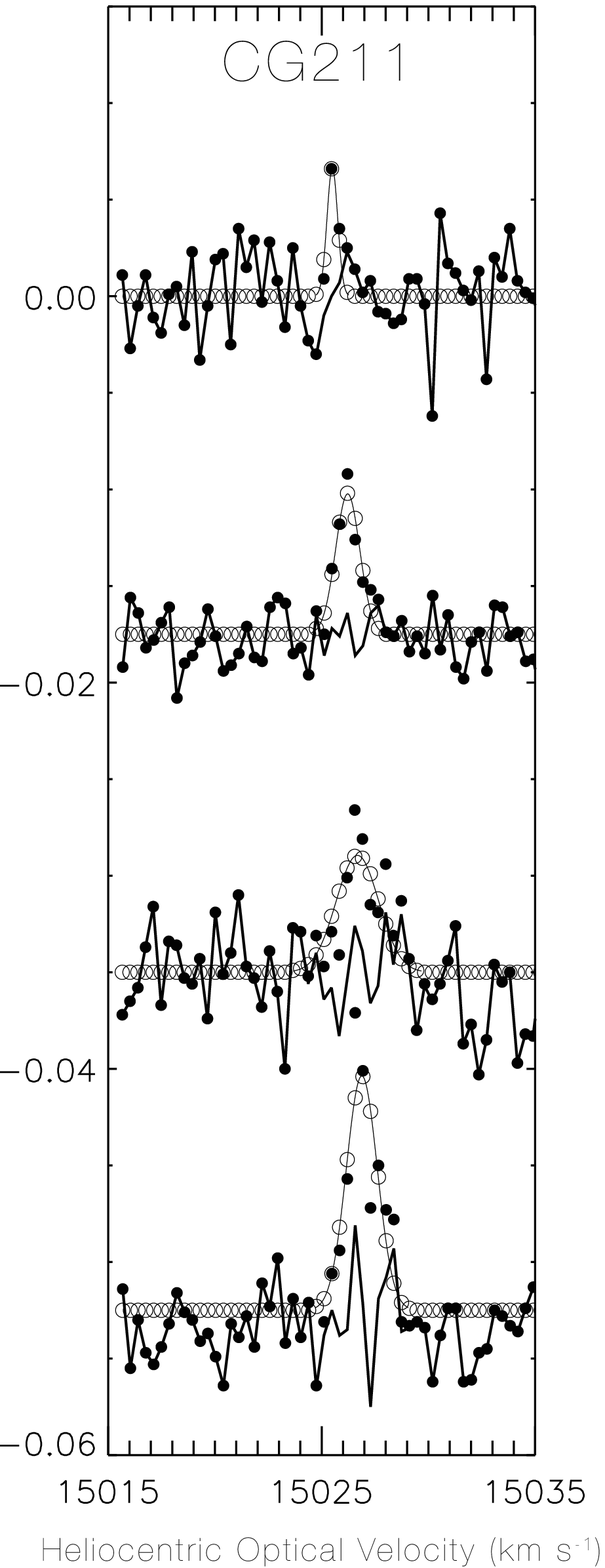}
\caption{Spectra of low-velocity features in CG\,211 obtained
with the Green Bank Telescope and the results of Gaussian component decomposition
(also see Table \ref{gauss_fit_table}). A wide ($\sim25$\,km\,s$^{-1}$)
stationary Gaussian component at $\sim14951$\,km\,s$^{-1}$ represents a broad
low-velocity plateau present at each epoch.\label{gauss_fit1}}
\end{figure}

\begin{figure}[p]
\plottwo{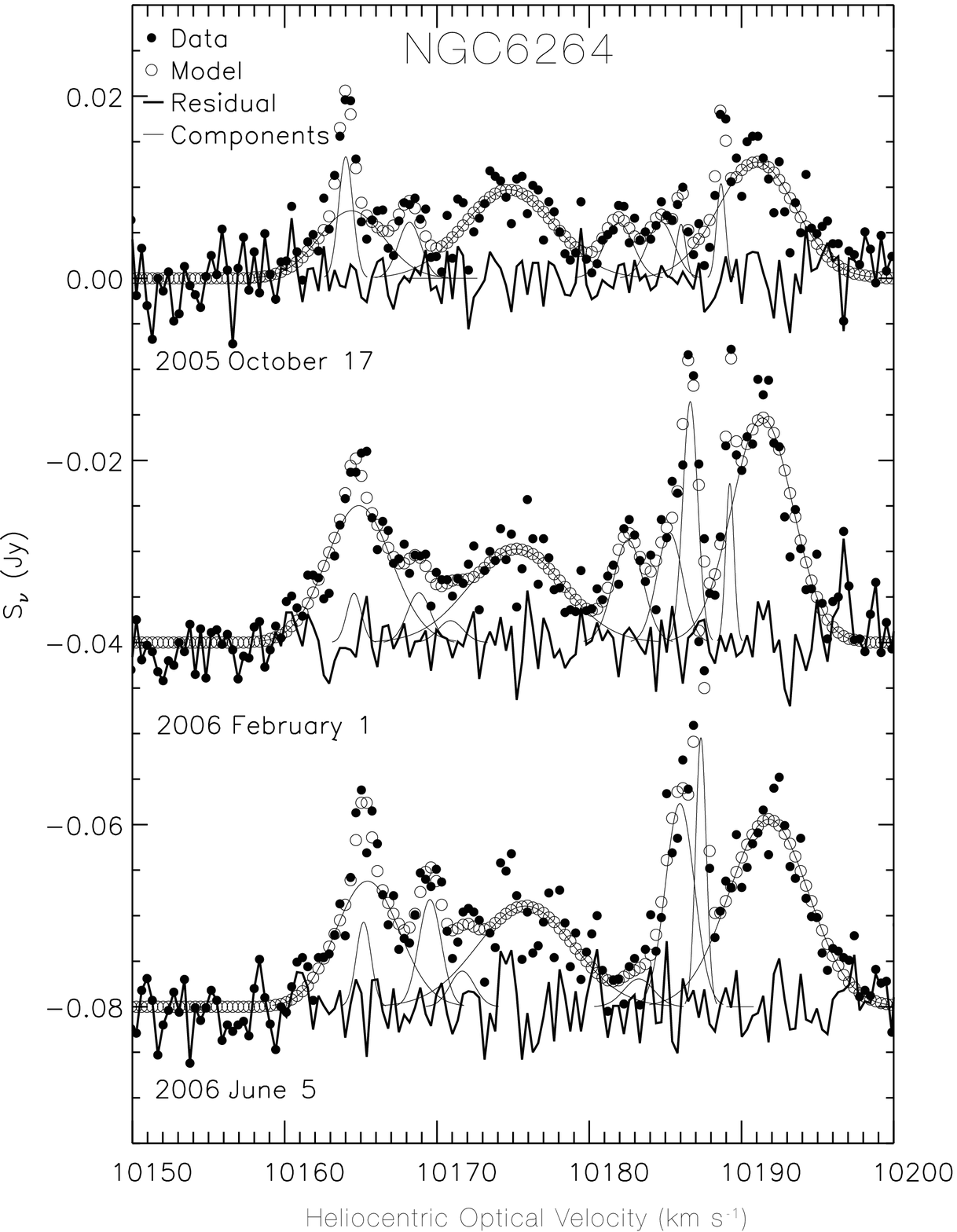}{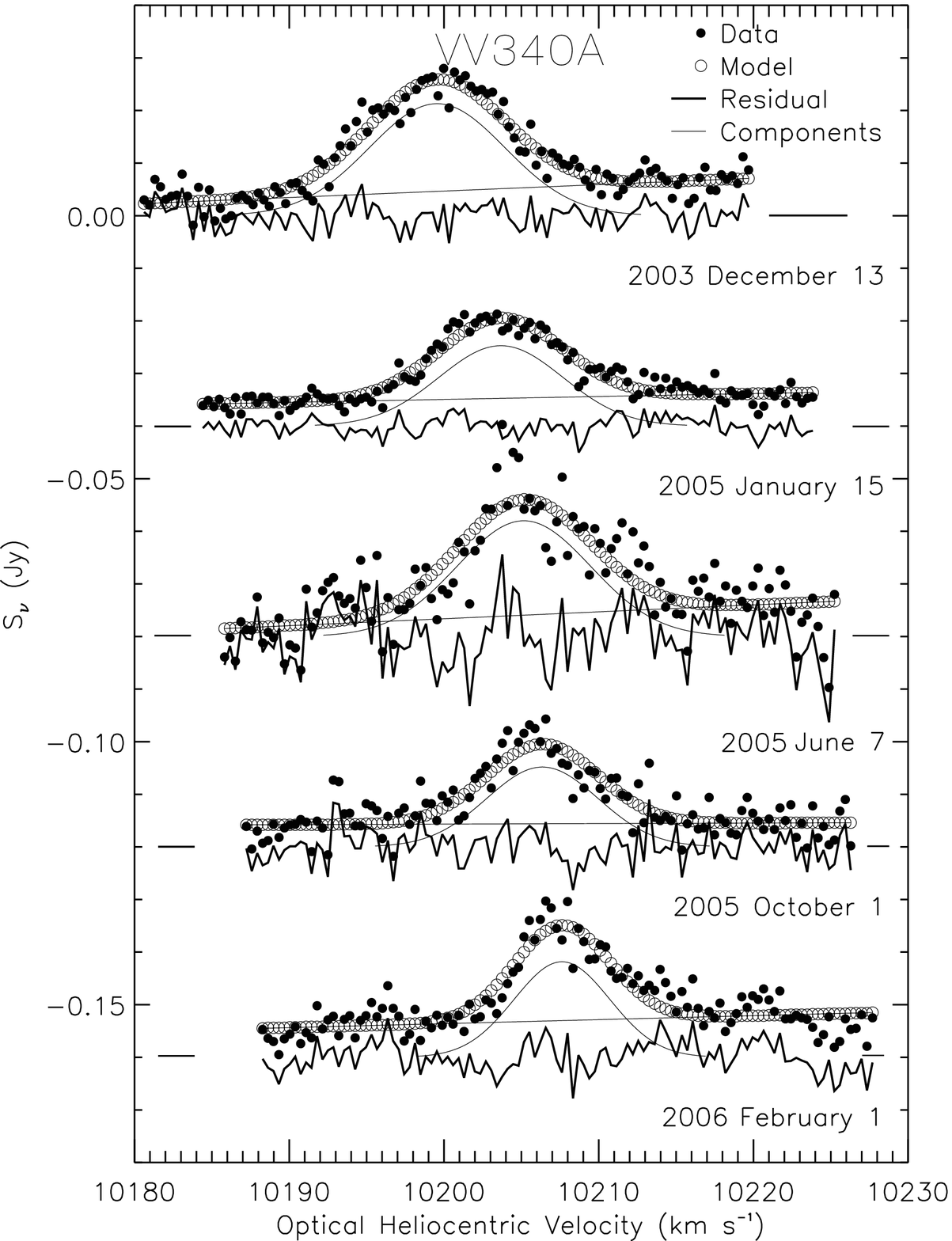}
\caption{Spectra of low-velocity features in NGC\,6264 {\it (left)} and
VV\,340A {\it (right)} obtained with the Green Bank Telescope and the
results of Gaussian component decomposition (also see Table~\ref{gauss_fit_table}). In the case of VV\,340A, horizontal lines to the left and right of the spectra mark the zero level, which is distinguished from the low-level broad emission component.
\label{gauss_fit2}}
\end{figure}

\begin{figure}[p]
\plotone{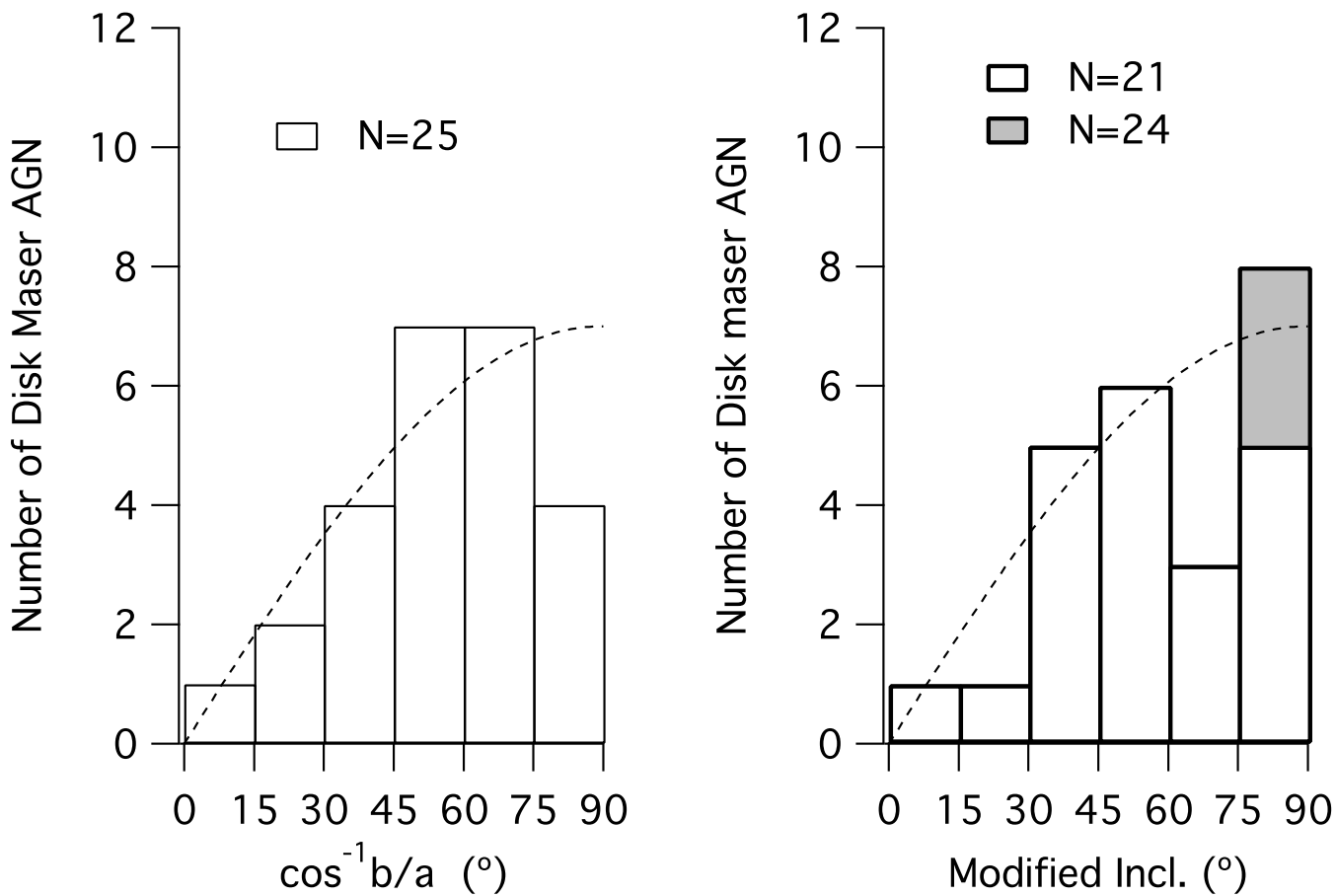}
\caption{Histograms of  inclinations for disk maser candidate host galaxies (Table~\ref{tab.incl}), demonstrating misalignment with the edge-on reference planes defined by disk maser candidates on sub-parsec scales. {(\it left}) Inclination computed from catalogued axial ratios at $B=25$ magnitude sec$^{-2}$  \citep{Vaucouleurs1991}, for published disk maser candidates that lie in spiral hosts.  The dashed line represents the expectation for a sample of randomly oriented galaxies, scaled to seven objects in the highest bin. ({\it right})  Inclination  computed with correction for  distortion of isophotes due to  galactic bulges \citep{fouque90}, using estimates of morphological type $T$ \citep{Vaucouleurs1991, Malkan1998, lauberts89}. Shaded bars represent three galaxies with inclinations very close to $90^\circ$ and for which a $\sim 1\sigma$ or $+0.6$ adjustment to  $T$ (e.g., Sa$\rightarrow$Sab)  was required to obtain allowable values of $\cos i$.}
\label{incl.histo}
\end{figure}

In a survey with the GBT of 40 inclined AGN, we have detected three new water
maser sources (NGC\,1320, NGC\,17, and IRAS 16288-3929; Figure~\ref{new.masers}). We classify two of these (NGC\,1320 and NGC\,17) as disk maser candidates based on their spectra. Three more cases of disk maser candidates were discovered through deep integrations toward known masers 
(NGC\,449, NGC\,2979, and NGC\,3735; Figure~\ref{other1}).  One previously marginal case was  confirmed (UGC\,4203; Figure~\ref{other2}).  
Inferred rotation speeds are $\sim 100$--300\,\kms. At present, the details of interpretation are sensitive to estimates of systemic velocity at optical wavelengths and  21~cm (the HI line).  These show significant scatter for some galaxies, and in general, comprehensive study is mandatory for each new maser detection to resolve uncertainties in classification. Interferometric follow-up of discoveries reported here will enable confirmation studies of disk geometry (e.g., radii, warping) and estimation of central engine mass, at least  for the sources with large numbers of maser Doppler components.  In general, systematic velocities that result from disk models may provide the best estimates of systemic velocity, and these might usefully be compared to the results of pointed optical and molecular line studies that include dynamical modeling.

For three additional maser sources exhibiting high-velocity maser emission (CG\,211, NGC\,6264, VV\,340A), we report secular drift in multiple low-velocity features---a probable manifestation of centripetal 
acceleration due to disk rotation. Combined with rotation velocities inferred from spectra, these centripetal 
accelerations suggest central engine masses of a few times $10^7$\,M$_\odot$ and mean disk radii of a few 
times $10^{-1}$\,pc. Measurable accelerations of $\sim$ 1--3\,\kms\ yr$^{-1}$ and large recessional 
velocities ($>10^4$\,\kms) make these three sources particularly attractive for interferometric study in pursuit of robust estimates of geometric distances and H$_0$. 

As beacons of highly inclined structures in AGN, at radii of $\sim 0.1$--1\,pc,  disk masers also highlight  those galactic nuclei in which we can define a reference plane with high accuracy and study (mis)alignment with surrounding kiloparsec-scale stellar disks. We find no preponderance of edge-on galactic disks among published disk maser candidate hosts.  
The comparison is in agreement with studies that infer orientations of accretion disks and geometrically thick tori from radio jets and ionization cones, though the maser data refer to a unique range of radii and probe 
inclination rather than position angle on the sky.

Our studies provide more candidates for the quest to measure H$_0$ with disk masers. The best available estimate of H$_0$ is uncertain by 5\% and achieved with a composite 
analysis of Cepheid and supernova data, with a zero point provided by the disk maser distance to NGC\,4258 \citep{riess09}. Ultimately, accuracy on the order of  1\% would match accuracies anticipated for CMB and other data sets in the near 
future and add a new independent constraint in cosmological parameter estimation.
However, achieving this accuracy for H$_0$ will be challenging, probably requiring detailed study of a disk maser sample at least three times larger
than what is available today, assuming maser distances with 10\% accuracy as the norm.  The minimum sample recessional velocity would 
be $10^4$\,\kms, though objects as nearby as 3200\,\kms\ may be useful where peculiar motions are known with uncertainties as small as 160\,\kms.  Difficulty in accurately estimating peculiar motions in some volumes may restrict this extended sample, e.g., near the Great Attractor or where 
uncertainty in Tully-Fisher or Fundamental Plane  distances are large.

\acknowledgments

The graduate dissertation by PK contributed to this work. 
We thank M. Elvis, E. Humphreys, and R. Narayan for helpful
discussions. This work benefitted from expert analysis 
of peculiar velocities and model flow fields offered by K. Masters.  
We thank M. Reid for discussions and the code used to fit 
the velocity drifts of maser features.  We appreciate support and 
assistance with GBT observing by J. Braatz, as well as overall encouragement.  
We thank C. Bignell for telescope scheduling that enabled the dissertation work 
and R. Maddalena for in-depth discussion of calibration.  This research has made extensive use of the NASA/IPAC Extragalactic Database (NED), which is operated by the Jet Propulsion Laboratory
(JPL). This work was supported by GBT student support program, grants GSSP004-0005 and GSSP004-0011, and in part by grant NNG05GK24G  from NASA.  The results presented here were obtained through NRAO observing programs GBT04C-031, GBT05A-015, GBT06A-056  (PI: Kondratko), and AK629 at the VLA.

\bibliography{ms}

\clearpage

\clearpage

\clearpage

\clearpage

\clearpage

\clearpage

\clearpage

\clearpage

\begin{deluxetable}{llllrccc}
\tabletypesize{\scriptsize}
\tablewidth{6.0in}
\tablecaption{Inclined AGN Surveyed for H$_2$O Maser Emission} 
\tablehead{
     \colhead{Galaxy\tablenotemark{\rm a}}              &
     \colhead{Type}              &
     \colhead{$\alpha_{\rm 2000}$\tablenotemark{\rm b}}      &
     \colhead{$\delta_{\rm 2000}$\tablenotemark{\rm b}}   &
     \colhead{$v_{sys}$\tablenotemark{\rm c}}    &
     \colhead{Date}           &
     \colhead{$T_{\rm sys}$\tablenotemark{\rm d}}      &
     \colhead{$1\sigma$\tablenotemark{\rm e}}  \\
     \colhead{}             &
     \colhead{}             &
     \colhead{(hh~mm~ss)}     &
     \colhead{(dd~mm~ss)}     &
     \colhead{(km s$^{-1}$)}       &
     \colhead{}        &
     \colhead{(K)}          &
     \colhead{(mJy)} 
}

\startdata

UGC\,12915 & LINER & 00~01~41.94 & $+$23~29~44.5 &          4336 & 2005-02-02 &           35 & 5.6  \\
UGC\,00050 & LINER & 00~06~40.15 & $+$26~09~16.2 &          7552 & 2005-02-01 &           34 & 6.1  \\
NGC\,17$\bigstar$
                      &   Sy2   & 00~11~06.55 & $-$12~06~26.3  &          5931 & 2006-04-18 &           47 & 3.2  \\
UGC\,00238 & LINER & 00~25~03.35 & $+$31~20~42.6 &          6796 & 2005-02-01 &          35 & 5.6  \\
M31                & LINER & 00~42~44.32 & $+$41~16~08.5 &          -300 & 2005-02-01  &          43 & 7.9  \\
IRAS\,01189+2156 & Sy2 & 01~21~44.27 & $+$22~12~35.1 &17478 & 2005-02-01  &          35 & 5.3  \\
UM\,319        & Sy2      & 01~23~21.18 & $-$01~58~36.0  &          4835 & 2005-09-21  &          56 & 9.2  \\
UGC\,00987 & Sy2      & 01~25~31.46 & $+$32~08~11.4 &          4658 & 2005-01-28 &           31 & 5.2  \\
NGC\,0660   & LINER & 01~43~01.70 & $+$13~38~34.0 &           850  & 2005-02-02 &            43 & 8.7  \\
UGC\,01282 & Sy2     & 01~49~29.87 & $+$12~30~32.6 &          5221 & 2005-02-02 &            38 & 6.5  \\
UGC\,01479 & Sy2     & 02~00~19.06 & $+$24~28~25.3 &          4927 & 2005-02-02 &            36 & 5.7  \\
UGC\,01757 & Sy2     & 02~17~23.05 & $+$38~24~49.9 &          5254 & 2005-01-28 &            30 & 5.5  \\
UGC\,02456 & Sy2     & 02~59~58.59 & $+$36~49~14.3 &          3605 & 2005-02-01 &            42 & 5.9  \\
UGC\,02638 & LINER & 03~17~02.21 & $+$01~15~17.9 &          7123 & 2005-09-21 &           54 & 10  \\
NGC\,1320$\bigstar$
                       & Sy2      & 03~24~48.70 & $-$03~02~32.2  &         2663  & 2005-09-30 &            47 & 5.9  \\
MCG\,-02-09-040 & Sy2 & 03~25~04.94 & $-$12~18~27.8 &       4495 & 2005-09-30 &           48 & 6.0  \\
SBS\,0811+584 &Sy2 & 08~16~01.30 & $+$58~20~01.0 &          7584 & 2005-09-30 &           43 & 5.6  \\
NGC\,2683   & Sy2       & 08~52~41.42 & $+$33~25~13.7 &           411 & 2005-09-30 &           44 & 6.5  \\
NGC\,3190   & LINER  & 10~18~05.77 & $+$21~49~55.8 &         1271 & 2005-10-01 &           46 & 5.9  \\
UGC\,05613 & LINER & 10~23~32.50 & $+$52~20~30.0 &          9620 & 2005-02-01 &           34 & 6.0  \\
NGC\,3593   & Sy2      & 11~14~37.00 & $+$12~49~04.0 &           628 & 2005-10-01  &           47 & 6.5  \\
NGC\,3628   & LINER & 11~20~16.95 & $+$13~35~20.1 &           843 & 2005-02-01  &            52 & 8.1 \\
NGC\,3753   & AGN    & 11~37~53.90 & $+$21~58~53.0 &         8727 & 2005-02-01  &            40 & 5.5  \\
MCG\,+03-38-017 & AGN & 14~47~53.39 & $+$19~04~37.1&12506& 2005-02-01     &          35 & 6.0 \\
IRASF\,15033+2617 & Sy2 & 15~05~27.93 & $+$26~05~29.3 &16579 & 2005-02-01 &         35 & 5.9  \\
IRASF\,15588+6522 &AGN&15~59~18.95 & $+$65~13~57.8 & 8975 & 2005-02-01   &          34 & 6.0  \\
IRAS\,16288+3929$\bigstar$
                        & Sy2      & 16~30~32.66 & $+$39~23~03.2  &          9161  & 2006-04-10  &        39 & 2.8  \\
UGC\,10593 & Sy2      & 16~52~18.87 & $+$55~54~19.8 &          8739 & 2005-09-21 &           70 & 14  \\
UGC\,10695 & LINER & 17~05~05.56 & $+$43~02~35.1 &          8328 & 2005-09-28 &           51 & 6.6  \\
NGC\,6503   & LINER & 17~49~27.11 & $+$70~08~39.6 &               60 & 2005-09-28 &           56 & 6.5  \\
MCG\,+11-22-046 & Sy2 & 18~22~02.20 & $+$66~36~37.7 &  4393 & 2005-09-28 &              57 & 7.0  \\
CGCG\,341-006 & Sy2 & 18~45~26.23 & $+$72~11~01.7 &      13880 & 2005-02-01 &           35 & 6.0  \\
NGC\,7013   & LINER & 21~03~33.31 & $+$29~53~49.3 &              779 & 2005-02-02 &          38 & 5.9 \\
IC\,1417        & Sy2      & 22~00~21.61 & $-$13~08~49.1 &          5345 & 2005-09-30 &             53 & 6.4  \\
NGC\,7331   & LINER & 22~37~04.09 & $+$34~24~56.3 &           816 & 2005-02-01 &             42 & 7.2  \\
UGC\,12201 & LINER & 22~49~09.55 & $+$34~59~30.5 &          5055 & 2005-02-02 &           34 & 4.9 \\
NGC\,7466   & Sy2      & 23~02~03.42 & $+$27~03~09.5 &          7493 & 2005-02-01 &           32 & 5.1  \\
UGC\,12318 & Sy2      & 23~02~08.01 & $+$25~40~13.7 &          9533 & 2005-02-01 &           34 & 5.5  \\
NGC\,7549   & AGN    & 23~15~17.23 & $+$19~02~30.1 &          4736 & 2005-02-02 &            35 & 5.5 \\
UGC\,12519 & Sy2     & 23~20~02.75 & $+$15~57~10.6 &          4378 & 2005-02-01 &            38 & 5.6  \\
 
\enddata

\tablenotetext {\rm a} {Newly discovered H$_2$O maser emission hosts are indicated by a $\bigstar$.}

\tablenotetext{\rm b}{Optical positions used for GBT pointing.  Uncertainties are typically less than 
$2''$.   Newly discovered maser emission in NGC\,1320 and IRAS\,16288+3929 lies within $1\sigma$ of the catalog positions for the two nuclei. \\ NGC\,1320: $\alpha_{\rm 2000}=03^h24^m48\rlap{.}^s70\pm0\rlap{.}^s02; \delta_{\rm 2000}=-03^\circ02'32\rlap{.}''3\pm0\rlap{.}''3$ (maser; VLA); \\ 
\hbox{\hspace{0.55in}} $\alpha_{\rm 2000}=03^h24^m48\rlap{.}^s70\pm0\rlap{.}^s08; \delta_{\rm 2000}=-03^\circ02'32\rlap{.}''2\pm1\rlap{.}''3$ \citep[nucleus;][]{Skrutskie2006}. \\  
IRAS16288+3929:  $\alpha_{\rm 2000}=16^h30^m32\rlap{.}^s66\pm0\rlap{.}^s02; \delta_{\rm 2000}=39^\circ23'03\rlap{.}''1\pm0\rlap{.}''3$ (maser; VLA); \\ 
\hbox{\hspace{0.95in}}  $\alpha_{\rm 2000}=16^h30^m32\rlap{.}^s65\pm0\rlap{.}^s02; \delta_{\rm 2000}=39^\circ23'03\rlap{.}''13\pm0\rlap{.}''5$ \citep[nucleus;][]{sdss3}.}

\tablenotetext{\rm c}{Systemic velocity (optical heliocentric) used in tuning the GBT observing bands. 
Velocities obtained from the NED.  When multiple measurements are available, those with the smallest listed uncertainties were chosen, in most cases.}

\tablenotetext{\rm d}{Average system temperature. Temperatures (and RMS noise) during 2005 January and February integrations have
been increased by a conservative 30\% factor to correct for apparent miscalibration of the noise-source. The correction factor was determined
using a receiver temperature of 16.5\,K, 6\,K contribution due to spillover and microwave background emission, and archival average forecast
ground temperature and 22 GHz opacity.}

\tablenotetext{\rm e}{RMS noise in a $24.4$\,kHz spectral channel corrected for
atmospheric opacity (i.e., referenced to outside the atmosphere), for the dependence of
antenna gain on elevation, and for uncertainty in the noise source power level (see note d.
Calibration uncertainties are discussed in Section~2.}

\label{survey}
\end{deluxetable}

\begin{deluxetable}{llllrrrrl}
\tabletypesize{\scriptsize}
\tablewidth{0pt} 
\tablecaption{Deep Integrations on Known H$_2$O Maser Sources}
\tablehead{
     \colhead{Galaxy}              &
     \colhead{Type\tablenotemark{\rm a}}              &
     \colhead{$\alpha_{\rm 2000}$\tablenotemark{\rm b}}      &
     \colhead{$\delta_{\rm 2000}$\tablenotemark{\rm b}}   &
     \colhead{$v_{sys}$\tablenotemark{\rm c}}    &
     \colhead{Date}           &
     \colhead{$T_{\rm sys}$\tablenotemark{\rm d}}      &
     \colhead{$1\sigma$\tablenotemark{\rm e}}   \\
     \colhead{}             &
     \colhead{}             &
     \colhead{(hh~mm~ss)}     &
     \colhead{(dd~mm~ss)}     &
     \colhead{(km s$^{-1}$)} &
     \colhead{}             &
     \colhead{(K)} &
     \colhead{(mJy)} 
     }

\startdata

NGC\,449 (Mrk\,1) & Sy2  & 01~16~07.25 & $+$33~05~22.4 & 4780 & 2006-04-19 & 48 & 2.4 \\

NGC\,2979       & Sy2  & 09~43~08.65 & $-$10~23~00.0 & 2720 & 2006-02-09 & 37 & 2.0  \\

NGC\,3735       & Sy2  & 11~35~57.30 & $+$70~32~08.1 & 2696  & 2006-02-09 & 37 & 3.9  \\

UGC\,4203 (Mrk\,1210) & Sy2  & 08~04~05.86 & $+$05~06~49.8 & $4046$ & 2006-04-24 & 44 & 3.3  \\

Arp\,299 (NGC\,3690) & AGN &  11~28~32.20 & $+$58~33~44.0 & $3121$  & 2006-04-27 & 46 & 2.9  \\

NGC\,4293       & LINER & 12~21~12.82 & $+$18~22~57.4 & $893$ & 2006-04-10 & 38 & 2.8  \\

NGC\,0235A     & Sy2 & 00~42~52.81 & $-$23~32~27.8 & $6664$  & 2006-02-01 & 58 & 3.8  \\

NGC\,2824       & AGN  & 09~19~02.22 & $+$26~16~12.0 & $2760$ & 2006-01-27 & 35 & 2.5  \\

IRASF\,19370-0131 & Sy2 & 19~39~38.91 & $-$01~24~33.2 & $6188$ & 2006-04-19 & 48 & 3.7  \\
\enddata

\tablenotetext{\rm a}{Activity type from the NED, except for NGC\,0235A, which is
from Kondratko et~al.\ 2006a.}

\tablenotetext{\rm b}{GBT pointing positions adopted from Kondratko et~al.\ 2006a for
NGC\,0235A, NGC\,4293 ($\sigma=0\rlap{.}''3$), from Greenhill et~al.\ 2003b for
NGC\,2824, NGC\,2979, IRASF\,19370-0131 ($\sigma=0\rlap{.}''2$), and from
the NED for NGC\,449 ($\sigma=0\rlap{.}''28$), UGC\,4203
($\sigma=0\rlap{.}''75$),  Arp\,299 (NGC\,3690,
$\sigma=10''$), NGC\,3735 ($\sigma=1\rlap{.}''3$).}

\tablenotetext{\rm c}{Systemic velocity (optical heliocentric) obtained from the NED used in tuning the GBT observing bands. 
Measurements for individual galaxies follow (in km s$^{-1}$).
{\it NGC\,449 (Mrk\,1)}: 
$4750\pm 16$ (optical; Falco et al. 1999), 
$4774\pm 9$ \citep[optical;][]{keel96}, 
$4823\pm 6$ \citep[HI;][]{Bottinelli1990};
\nocite{falco1999}
{\it NGC\,2979}: 
$2664\pm5$ (HI; Saunders et al. 2000),
$2686\pm8$ \citep[HI;][]{Paturel2003},
$2720\pm15$ (optical; Fisher et al. 1995), 
$2731\pm26$ \citep[optical;][]{Fairall1992};
\nocite{Fisher1995, Saunders2000}
{\it NGC\,3735}: 
$2597\pm42$ \citep[optical;][]{Sandage1978},
$2684\pm39$ (optical; Falco et al. 1999),
$2696\pm6$ \citep[HI;][]{Staveley1988},
$2696\pm15$ \citep[HI;][]{Fisher1981},
$2738\pm39$ \citep[optical;][]{Vaucouleurs1979};
\nocite{falco1999}
{\it UGC\,04203 (Mrk\,1210)}: 
$3988\pm33$ \citep[optical;][]{Fisher1995},
$4046\pm 9$ \citep[HI;][]{Bottinelli1990}, 
$4065\pm 42$ (optical; Falco et al. 1999);
{\it Arp\,299 (NGC\,3690)}: see Hibbard \& Yun 1999 concerning nomenclature):
\nocite{Hibbard1999}
$3040\pm27$ \citep[optical;][]{Garcia2006}, 
$3080\pm40$ \citep[radio;][]{Zhao1997};
{\it Arp\,299 (IC\,694)}:
$3120\pm50$ \citep[radio;][]{Zhao1997},
$3121\pm27$ \citep[optical;][]{Garcia2006};
{\it NGC\,4293}: 
$717\pm 58$ \citep[optical;][]{Vaucouleurs1991},
$867\pm29$ (optical; J. Huchra, private communication); 
$883\pm 54$ (optical; Falco et al. 1999), 
$893\pm 8$ \citep[optical;][]{diNella1995}, 
$933\pm38$ \citep[optical;][]{Huchra1983};
{\it NGC\,0235A (ESO\,474-G016)}:
$6692\pm 37$ (optical; da Costa et al. 1998),
$6692\pm 28$ \citep[optical;][]{daCosta1991}, 
$6712\pm28$ (optical; Wegner et al. 2003),
$6765\pm26$ \citep[optical;][]{Jones2005};
\nocite{Wegner2003}
{\it NGC\,2824}: 
$2760\pm 23$ \citep[optical;][]{Huchra1983}, 
$2755\pm 44$ (optical; Falco et al. 1999);
{\it IRASF\,19370-0131}: 
$6188\pm 36$ \citep[optical;][]{Strauss1992},
$6262\pm 50$ \citep[optical;][]{Jones2005}.
}

\tablenotetext{\rm d}{Average system temperature (not corrected for opacity).}

\tablenotetext{\rm e}{RMS noise in a 24.4-kHz spectral channel corrected for
atmospheric opacity (typically from $0.02$ to $0.04$) and for the dependence of
antenna gain on elevation.}
\label{TableOther}
\end{deluxetable}

\begin{deluxetable}{llcc} 
\tabletypesize{\scriptsize}
\tablecaption{Monitoring Log for Secular Drifts}
\tablewidth{4in}
\tablehead{
     \colhead{Source}              &
     \colhead{Epoch}       &
     \colhead{Time Monitor\tablenotemark{\rm a}}  &
     \colhead{Mean Spectrum\tablenotemark{\rm b}}
     }

\startdata

CG\,211\tablenotemark{\rm c} & 2003-12-08 & x & ...\\ 
                & 2005-01-15 & x & x \\
                & 2005-09-28 & x & x \\
                & 2006-01-22 & x & x \\
NGC\,6264 & 2005-01-27 & ... & x \\
                     & 2005-10-17 & x & x \\
                     & 2006-02-01 & x & x \\
                     & 2006-06-05 & x & ... \\
VV\,340A & 2003-12-13 & x & ... \\
                  & 2005-01-15 & x & x \\
                  & 2005-06-07 & x & ... \\
                  & 2005-10-01 & x & x \\
                  & 2006-02-01 & x & x \\
\enddata
\tablenotetext{\rm a}{Epochs over which spectral features were tracked, enabling estimation of accelerations (Figures 6,7).}
\tablenotetext{\rm b}{Epochs used to generate the weighted mean spectrum (Figure 5). For each source, the best three spectra were selected, with particular attention to the flatness of the spectral baseline.}
\tablenotetext{\rm c}{The CG designation refers to the Case low-dispersion spectroscopic survey \citep[][and later papers by the same authors]{case83}.}
\label{accel.log} 
\end{deluxetable}

\begin{deluxetable}{lccccc}
\tabletypesize{\scriptsize}
\tablewidth{5in}
\tablecaption{Secular Drifts in Low-Velocity (Systemic) Emission} 
\tablehead{
     \colhead{Source}              &
     \colhead{Epoch\tablenotemark{\rm a}}      &
     \colhead{$\chi^2_R$\tablenotemark{\rm b}}      &
     \colhead{Velocity\tablenotemark{\rm c}}   &
     \colhead{Drift\tablenotemark{\rm d}}  &
     \colhead{Epochs}  \\
     \colhead{}             &
     \colhead{}             &
     \colhead{}             &
     \colhead{(km s$^{-1}$)} &
     \colhead{(km s$^{-1}$ yr$^{-1}$)}
}

\startdata

CG\,211\tablenotemark{\rm e} & 2003 Dec 08 & $1.7$ & $14940.92\pm 0.06$ & $0.405\pm 0.005$ & 4\\
             &              &       & $14943.2\pm 0.4$ & $0.90\pm 0.10$ \\
             &              &       & $14951.9\pm 0.1$ & $0.93\pm 0.03$ \\
             &              &       & $14954.70\pm 0.06$& $0.58\pm 0.03$ \\
             &              &       & $14963.9\pm 0.1$& $0.85\pm 0.07$ \\
             &              &       & $14967.7\pm 0.1$& $1.55\pm 0.04$ \\
             &              &       & $14969.9\pm 0.1$& $1.88\pm 0.07$ \\
             &              &       & $14976.0\pm 0.2$& $0.11\pm 0.05$ \\
             &              &       & $14949.2\pm 0.9$& $1.45\pm 0.09$ \\
             & 2003 Dec 08  & $1.0$ & $15025.5\pm 0.1$ &  $0.65\pm 0.04$ \\
             &              &       &                                   & (avg. = 0.9)\tablenotemark{\rm f} \\
NGC\,6264    & 2005 Oct 17   & $1.2$ & $10164.0\pm 0.2$ & $1.9\pm0.2$ & 3 \\
             &              &       & $10164.4\pm 0.3$ & $1.7\pm0.5$  \\
             &              &       & $10168.2\pm 0.3$ & $2.1\pm0.3$ \\
             &              &       & $10174.7\pm 0.3$ & $1.7\pm0.6$ \\
             &              &       & $10181.9\pm 0.3$ & $2.1\pm0.6$ \\
             &              &       & $10184.7\pm 0.4$ & $2.0\pm0.4$ \\
             &              &       & $10186.0\pm 0.1$ & $2.1\pm0.1$ \\
             &              &       & $10188.7\pm 0.2$ & $2.0\pm0.2$ \\
             &              &       & $10191.0\pm 0.2$ & $1.2\pm0.4$ \\
             &              &       &                                   &(avg. = 2)\tablenotemark{\rm f} \\
VV\,340A\tablenotemark{\rm e}  & 2003 Dec 13& $1.4$ & $10199.6\pm 0.1$ & $3.8\pm0.1$ & 5 \\

\enddata

\tablenotetext{\rm a}{Reference epoch for component velocities.}

\tablenotetext{\rm b}{Reduced chi-squared for the least-squares decomposition of line profiles into Gaussian components, in which reference-epoch velocities, secular drifts, time-varying widths, and light curves are simultaneously fit for each Gaussian component using a time-series of spectra.  Only components with signal-to-noise  $>3$ for two or more epochs are reported.}

\tablenotetext{\rm c}{Reference epoch for fitted velocity.} 

\tablenotetext{\rm d}{Velocity drift of a Gaussian component.}

\tablenotetext{\rm e} {{Low-velocity} plateaus were modeled by an FWHM $\sim 25$\,\kms\ stationary Gaussian component for  CG\,211 and a first-order polynomial for VV\,340A. These components are not reported here.}

\tablenotetext{\rm f}{Unweighted arithematic average for use in the calculation of entries in Table~5 (see discussion in Section 4.1).}

\label{gauss_fit_table}
\end{deluxetable}

\begin{deluxetable}{lcc}
\tabletypesize{\scriptsize}
\tablewidth{2.0in}
\tablecaption{Estimate of Enclosed Mass and Disk Radii}
\tablehead{ 
  \colhead{Galaxy} &
  \colhead{Radius\tablenotemark{a}} &
  \colhead{Mass\tablenotemark{a}} \\
  \colhead{} &
  \colhead{(pc)} &
  \colhead{$(10^7$M$_\odot$)}
}
\startdata
CG\,211     & 0.4 & 3 \\
NGC\,6264 & 0.2 & 2 \\
VV\,340A   & 0.3 & 7 \\ 
\enddata

\tablenotetext{a} {Because of the uncertainties due to the lack of VLBI data as discussed in Section 4.1, the accuracies of the masses and radii may be as large as half an order of magnitude.}

\label{mass.radius}
\end{deluxetable}

\begin{deluxetable}{lccccc}
\tabletypesize{\scriptsize}
\tablewidth{4in}
\tablecaption{Inclinations of Galaxies that have Candidate or Confirmed Disk Masers} 
\tablehead{
     & & & & \multicolumn{2}{c}{{Evidence for Disk Maser}\tablenotemark{\rm e} } \\
     \colhead{Source\tablenotemark{\rm a} }              &
     \colhead{$T$\tablenotemark{\rm b} }      &
     \colhead{$\cos^{-1}(b/a)$\tablenotemark{\rm c} }      &
     \colhead{$i_{mod}$\tablenotemark{\rm d} }  &
     \colhead{Spectrum}  &
     \colhead{Map}  \\
     \colhead{}             &
     \colhead{}             &
     \colhead{($^\circ$)}     &
     \colhead{($^\circ$)}    &
     \colhead{Ref.}     & 
     \colhead{Ref.}      \\ 
}

\startdata
NGC\,4945\tablenotemark{\rm f} & 6 & 79 & 86 & 1 & 1 \\
Circinus\tablenotemark{\rm f}       & 3 & 64 & 69 & 2, 3 & 16 \\
NGC\,4258\tablenotemark{\rm f} & 4 & 67 & 71  & 4  & 17 \\
NGC\,1386\tablenotemark{\rm f} & -0.6 & 68 & (87) & 5 & 18 \\
NGC\,1068\tablenotemark{\rm f} & 3 & 32 & 33 & 6  & 19 \\
NGC\,3079\tablenotemark{\rm f}  & 7 & 80 & 85 & 7  & 20 \\
NGC\,4388  & 3 & 77 & (87) & 8 &  ... \\
NGC\,3735 & 5 & 78 & (86) & 9  & ... \\
NGC\,1320 & 1 & 70 & 85 & 9  & ... \\
NGC\,5728 & 1 & 54 & 60 & 8 & ... \\
IC 2560\tablenotemark{\rm f} & 3.3 & 51 & 53 & 5 & 21 \\
NGC\.5793 & 3 & 70 & 77 & 10 & ... \\
NGC\,3393\tablenotemark{\rm f} & 1 & 24 & 26 & 11 & 22 \\
NGC\,591 & 0 & 34 & 37 & 8 & ... \\
Mrk\,1419 & 1 & 46 & 50 & 12  & ... \\
NGC\,6926 & 4 & 46 & 48 & 13 & ... \\
NGC\,5495 & 4.6 & 37 & 38 & 11 & ... \\
NGC\,6323\tablenotemark{\rm f} & 2 & 71 & 81 & 8 & 23 \\
NGC\,2979 & 1 & 51 & 55 & 9 & ... \\
UGC\,3789\tablenotemark{\rm f} & 2 & 29 & 31 & 14 & 24 \\
UGC\,3193 & 2.3 & 68 & 74 & 14 & ... \\
UGC\,4203  & 1 & 0 & 0 & 9 & ... \\
NGC\,449 & 5 & 52 & 54 & 9 & ... \\
ESO\,269-G012 & --2 & 39 & 41 & 13 & ... \\
NGC\,6264 & ... & 51 & ... & 15 & ... \\
\enddata

\tablenotetext{\rm a}{Published disk maser candidates,in order of increasing recessional velocity. Missing isophotal sizes or merger activity exclude IRASF 22265-1826; CG\,211, VV\,340A,  3C\,403, NGC\,17.}
\tablenotetext{\rm b}{Morphological classification.}
\tablenotetext{\rm c}{Inclination from axial ratio $b/a$.}
\tablenotetext{\rm d}{Modified inclination \citep{fouque90}. Numbers in parentheses are obtained by adding $1\sigma$ to $T$ to obtain a real inclinations.}

\tablenotetext{\rm e}{References for identification of disk maser candidates via spectroscopy: 
(1) Greenhill et al. (1997c)\nocite{Greenhill1997c}; (2) \citet{nakai95}; (3) \citet{greenhill97a};  
(4) \citet{nakai93}; (5) \citet{Braatz1996}; (6) \citet{cl86}; (7)  \citet{hagiwara02};
(8)  \citet{Braatz2004}; (9) this work; (10)  \citet{hagiwara97}; (11) \citet{Kondratko2006a};
(12) \citet{henkel02};  (13)  \citet{Greenhill2003survey}; (14)  \citet{bg08}; (15)  \citet{Kondratko2006b}. 
References for confirmation of disk maser identification via VLBI study: (16) \citet{Greenhill2003}; 
(17) \citet{Argon2007}; (18) \citet{Braatz1997a}; (19) \citet{Greenhill1997}; (20) \citet{Kondratko2005}; 
(21) \citet{Ishihara2001}; (22) \citet{kondratko08}; (23) \citet{Braatz2007}; (24) \citet{reid09}. }

\tablenotetext{\rm f}{Disk maser status confirmed by VLBI observations.}

\label{tab.incl}
\end{deluxetable}

\end{document}